\documentclass[pdflatex,sn-mathphys-num]{sn-jnl}% Math and Physical Sciences Numbered Reference Style
\usepackage{graphicx}%
\usepackage{multirow}%
\usepackage{amsmath,amssymb,amsfonts}%
\usepackage{amsthm}%
\usepackage{mathrsfs}%
\usepackage[title]{appendix}%
\usepackage{xcolor}%
\usepackage{textcomp}%
\usepackage{manyfoot}%
\usepackage{placeins}
\usepackage{lineno}
\newcommand{\dpho}{\ensuremath{A^{\prime}}}

\newcommand{\egev}{\ensuremath{\,\textrm{GeV}}}

\newcommand{\nc}{\newcommand}
\def\theequation{\thesection.\arabic{equation}}
\nc{\beq}{\begin{equation}}
\nc{\eeq}{\end{equation}}
\nc{\barray}{\begin{eqnarray}}
\nc{\earray}{\end{eqnarray}}
\nc{\barrayn}{\begin{eqnarray*}}
\nc{\earrayn}{\end{eqnarray*}}
\nc{\bcenter}{\begin{center}}
\nc{\ecenter}{\end{center}}
\nc{\mc}{\mathcal}
\nc{\er}[1]{(\ref{eq:#1})}
\nc{\onehalf}{\frac{1}{2}}
\nc{\partialbar}{\bar{\partial}}
\nc{\psit}{\widetilde{\psi}}
\nc{\Tr}{\mbox{Tr}}
\nc{\hc}{\mbox{H.c.}}
\nc{\ev}{\;\mathrm{eV}}
\nc{\mev}{\;\mathrm{MeV}}
\nc{\gev}{\;\mathrm{GeV}}
\nc{\tev}{\;\mathrm{TeV}}
\nc{\f}{\frac}

\begin{document}

\title[Article Title]{Sensitivity of the FCC-ee to the decay of
a dark photon  into a $\mu^+\mu^-$ pair}


\author*[1]{G. Polesello}\email{giacomo.polesello@cern.ch}

\affil[1]{INFN Sezione di Pavia, Via Bassi 6, 27100 Pavia, Italy}


\abstract{The production of a dark photon $\dpho$ at the
proposed CERN FCC-ee collider is investigated. The study addresses the associated
production $e^+e^-\rightarrow\gamma\dpho$ followed by the decay $\dpho\rightarrow\mu^+\mu^-$.
The 95\% CL sensitivity on the mixing between the photon and the dark photon
is evaluated in the $m_{\dpho}$ mass range 0.4-360~GeV, based on a
parametrised simulation of the IDEA detector. The study is performed both for 
prompt and long-lived decays of the $\dpho$.}

\maketitle

\section{Introduction}\label{sec::introduction}
The next generation of high-energy particle colliders is under
active discussion in the particle physics community.
A very attractive option that is being discussed is $e^+e^-$ circular colliders, such as the CERN FCC-ee \cite{FCC:2018evy}. These machines will provide access to
a broad range of physics studies, from precision measurements
of the Higgs boson and of Standard Model (SM) parameters, to direct
searches for physics beyond the Standard Model (BSM).

An interesting group of BSM models are the dark sectors \cite{Alexander:2016aln, Battaglieri:2017aum}, which are well-motivated candidates to explain the observed abundance of dark matter.
In dark sector models the SM interacts with dark matter through a weakly coupled SM
neutral mediator. Different scenarios have been proposed, corresponding to different
types of mediators.  The Higgs, vector, and neutrino portals, where the mediator 
is, respectively, a scalar, a vector, and a fermion field have been the subject of
intense phenomenological investigation.

The present study aims at evaluating the sensitivity to the production of 
vector mediators at the FCC-ee. All of the FCC-ee runs are considered,
namely the $Z$-pole, $WW$, $ZH$ and $t\bar{t}$,
at  centre-of-mass  energies respectively of $\sqrt{s}=91.2,\, 160$,  $240$ and $365\egev$ and 
with target integrated luminosities respectively of 
$L_{int}= 205,\,19.2,\,10.8, 2.7~\mathrm{ab}^{-1}$. \cite{FCC:2025lpp} 
%the center-of-mass energies $\sqrt{s}=88, 91, 94\egev$, corresponding to the production
%of approximately $6\times10^{12}$ $Z$ bosons. 

We focus on a specific implementation of the vector portal, the so-called dark photon 
or hypercharge portal, as defined in \cite{Curtin:2014cca}.  The dark photon mass region 
ranging from $0.4$ to  $360 \egev$ is investigated for this model, based on the study of 
the process $e^+e^-\rightarrow \gamma A^{\prime}$ followed by the decay 
$A^{\prime}\rightarrow\mu^+\mu^-$.
The decay into muons is expected to provide better sensitivity than the decays
into electrons, taus or hadrons, because of lower backgrounds and more precise experimental
reconstruction of the final state, yielding better separation power of 
signal from background. A significant additional sensitivity is however expected
from the consideration of all of the \dpho~ decay modes, and detailed analyses
in this direction are a possible future development of the present study.

The phenomenology of the model and the status of the experimental searches are reviewed in 
detail in \cite{Fabbrichesi:2020wbt} and \cite{Graham:2021ggy}.
%The sensitivity to the model of various  future accelerators is studied in 
%\cite{Curtin:2014cca}, but that study for $e^+e^-$ concentrates on the exotic decay of the 
%Higgs boson into two dark photon. 
Phenomenological studies addressing the $\dpho\rightarrow\mu^+\mu^-$ signature 
at future $e^+e^-$ colliders are documented in \cite{He:2017zzr} and \cite{Airen:2024iiy}.

We extend and complete the previous work by assessing
the experimental reach of the FCC-ee based on a detailed parametrised  simulation of the proposed IDEA detector \cite{IDEAStudyGroup:2025gbt} 
and on the consideration of the leading backgrounds for all 
the FCC-ee runs. This approach provides a realistic
estimate of the discovery potential and it highlights the  dependence of the result from key elements of the detector performance.
In particular, the crucial resolution on the two-muon invariant mass is evaluated on the basis of the simulation of the muon track measurement in the tracker of the IDEA detector.
The complete kinematics of the events is used in order to separate the signal from the backgrounds, rather than only exploiting the mass constraints, yielding a significant improvement on the results of \cite{Airen:2024iiy}.

In addition, exploiting the excellent vertexing capabilities of the IDEA detector, a detailed experimental analysis targeting long-lived 
decays of the dark photon is developed. This analysis gives access to an area of the parameter space of the benchmark model which is outside of the reach of projected collider
and beam dump experiments, and was not considered in previous publications.

The paper is organised as follows: the effective dark photon model is first introduced
and briefly discussed.  In the following sections the Monte Carlo event generation is discussed, followed by a description of the adopted approach to detector simulation. 
On this basis, analyses targeting prompt and long-lived signatures are developed in the following sections. Finally, the additional coverage of FCC-ee  with respect to the projections for the main experiments expected to take data
in the coming years is assessed for the studied benchmark model.

\section{The model and its parameters}\label{sec:model}
The benchmark model addressed in the present study is defined in \cite{Curtin:2014cca}. 
It implements kinetic mixing between a
broken dark Abelian gauge symmetry, $U(1)_D$, and the SM hypercharge,
$U(1)_Y$.  The relevant gauge terms in the
Lagrangian are
\beq\label{eq:KM}
\mathcal{L} \subset -\frac{1}{4} \,\hat B_{\mu\nu}\, \hat B^{\mu\nu} - \frac{1}{4} \,\hat Z_{D\mu\nu}\, \hat Z_D^{\mu\nu}  + \f{1}{2}\,\f{\epsilon}{\cos\theta} \,\hat Z_ {D\mu\nu}\,\hat B^{\mu\nu} + \f{1}{2}\, m_{D,0}^2\, \hat Z_D^\mu \, \hat Z_{D\mu}\, .
\eeq
Here the hatted fields indicate the original fields with non-canonical
kinetic terms, before any field redefinitions. The $U(1)_Y$ and
$U(1)_D$ field strengths are respectively $\hat B_{\mu\nu}
=\partial_\mu \hat B_{\nu} - \partial_\nu \hat B_{\mu}$ and $\hat
Z_{D\mu\nu} =\partial_\mu \hat Z_{D\nu} - \partial_\nu \hat Z_{D\mu}$,
$\theta$ is the Weinberg mixing angle, and $\epsilon$ is the kinetic
mixing parameter. After electroweak symmetry breaking three neutral
gauge bosons are present in the model: the photon, the $Z$, and a 
dark photon, which will be called \dpho~ in the following.

\begin{figure}[h] \centering
\includegraphics[width=0.5\textwidth]{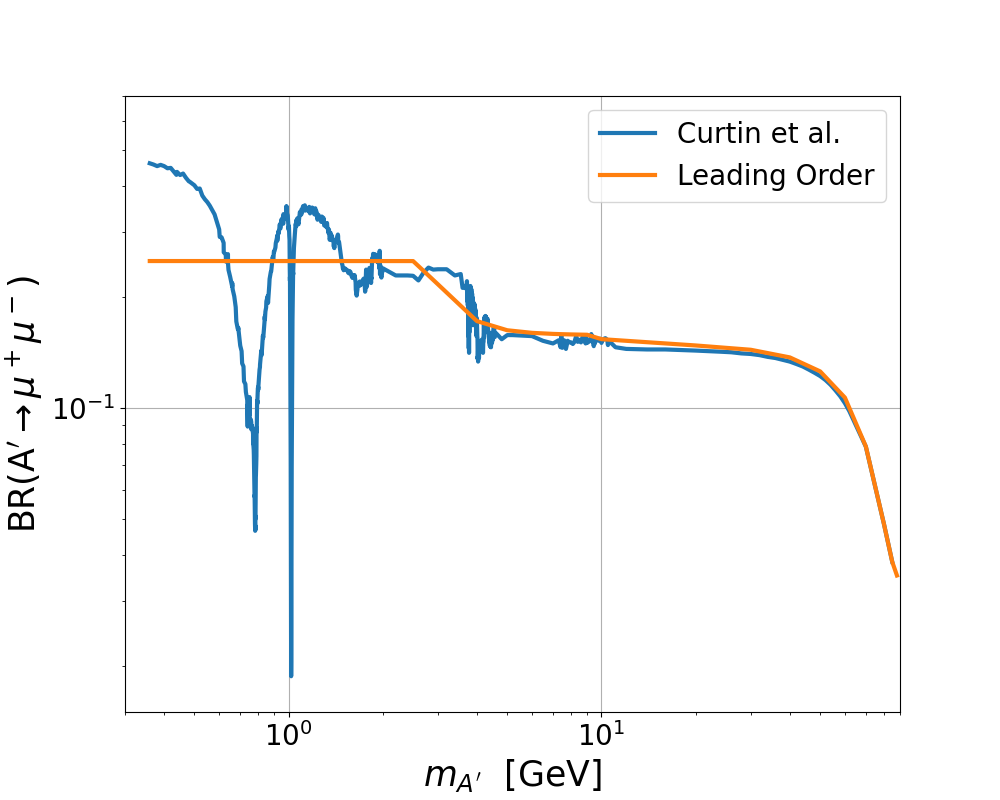}
\caption{Value of $BR\rightarrow\mu^+\mu^-$  as calculated
in \cite{Curtin:2014cca} (blue) and leading order calculation as implemented
in MG5aMC@NLO  (orange) as a function of $m_{\dpho}$.}\label{fig:brdph}
\end{figure}

%Both the production cross-section and the decay width of the dark photon are
%proportional to $\epsilon^2$. 
The leading order couplings of the dark photon to fermions
are calculated in \cite{Curtin:2014cca}, they are proportional to $\epsilon$ 
and depend on the ratio of $m_{\dpho}$ and $m_Z$. 
For $\epsilon\ll 1$ and $m_{\dpho}\ll m_Z$ the couplings are photon-like, 
and they evolve towards $Z$-like as $m_{\dpho}$ approaches $m_Z$.
The leading order calculation does not provide
an accurate prediction of the  dark photon width and its branching fraction 
into muons for $m_{\dpho}<10$~GeV. 
It is in fact necessary to account for the hadronic resonances into which 
the dark photon dominantly 
decays if its mass is close to the resonance mass.
A detailed calculation is provided in \cite{Curtin:2014cca},
together with a computer-readable table of widths and branching fractions 
for $m_{\dpho}$ in the range 0.36-88~GeV, which will be used throughout 
this paper. A comparison of $BR(\dpho\rightarrow\mu^+\mu^-)$ as calculated 
in \cite{Curtin:2014cca} with the leading order one, as  implemented 
by the same group in the \verb+MG5aMC@NLO+ generator, is
shown in Figure~\ref{fig:brdph}. The latter gives a constant value of 0.25 
for $m_{\dpho}<3$~GeV, whereas the full calculation exhibits a complex structure, 
with large dips for masses corresponding to the masses of the $\omega$ (782.6 MeV) 
and $\phi$ (1020 MeV) resonances.

\begin{figure}[h] \centering
\includegraphics[width=0.49\textwidth]{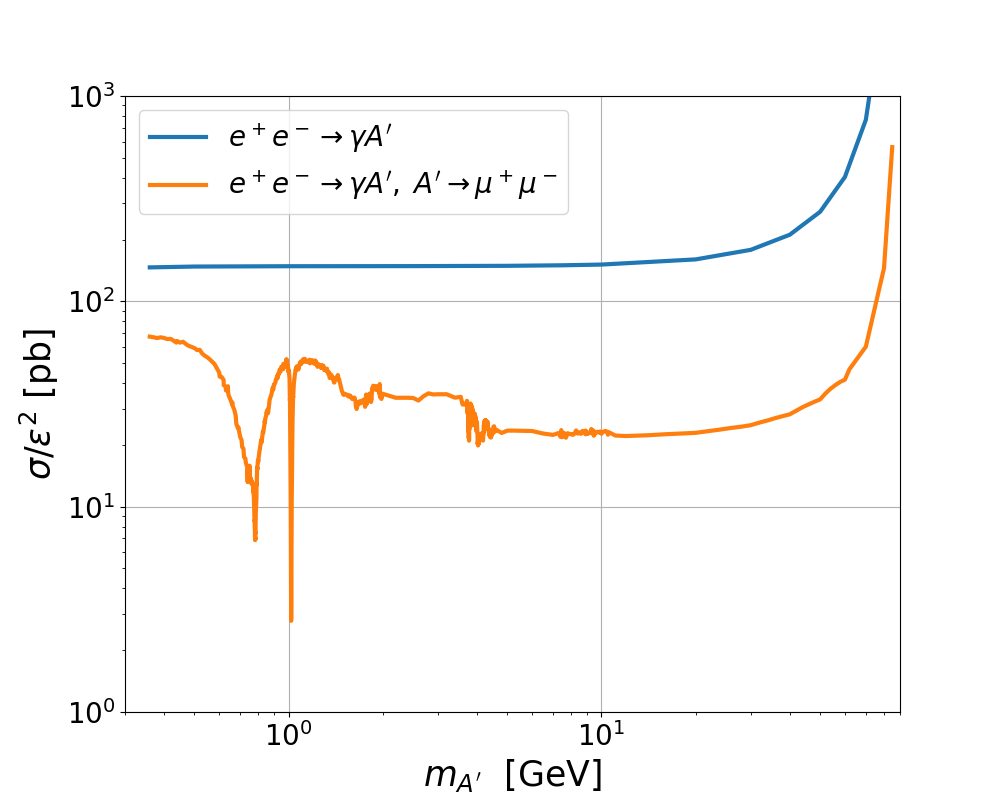}
\includegraphics[width=0.49\textwidth]{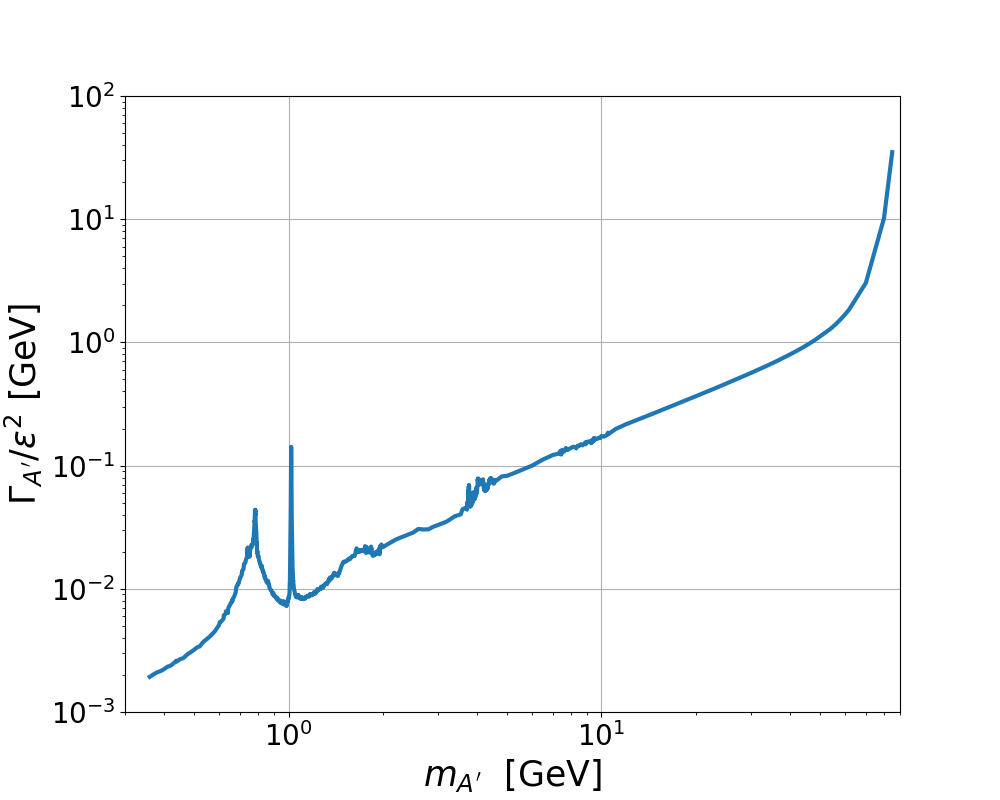}
\caption{Left: Cross-section  for \dpho~ production in pb  divided by $\epsilon^2$
as a function of $m_{\dpho}$. Right: Width in GeV of the $A^{\prime}$ divided by $\epsilon^2$ as a function of $m_{\dpho}$.}\label{fig:dphoxs}
\end{figure}

The resulting values of the \dpho~ cross-section for $\sqrt{s}=91.2$~GeV and total width 
$\Gamma_{\dpho}$ are proportional to $\epsilon^2$, and 
are shown in Figure~\ref{fig:dphoxs} for $\epsilon=1$. 

\begin{figure}[h] \centering
\includegraphics[width=0.7\textwidth]{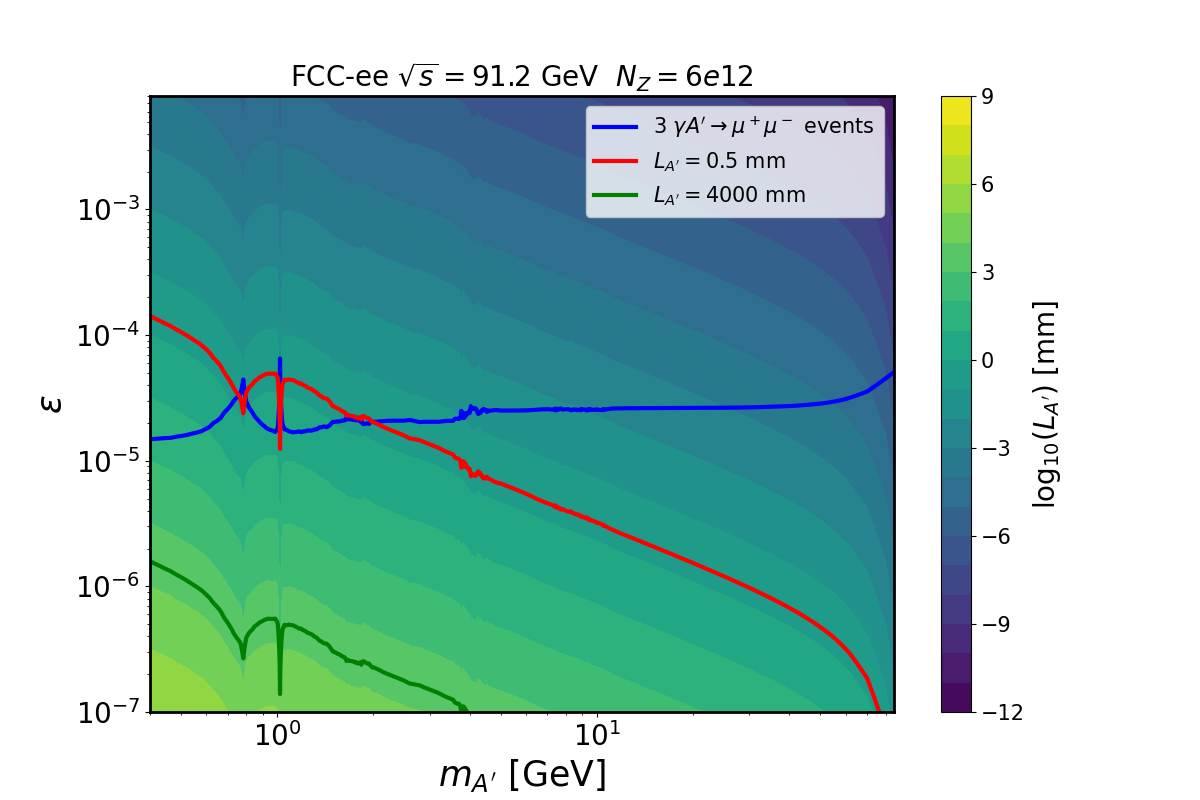}
\caption{$L_{\dpho}$ in the $m_{\dpho}-\epsilon$ plane calculated for
the process $e^+e^-\rightarrow\gamma\dpho$  at $\sqrt{s}=91.2$~GeV. The red and 
the green line bound the area in the parameter space where a long-lived 
\dpho~ search can be performed, and the blue line shows the parameters
corresponding to the production of 3 events for the expected FCC-ee 
	statistics of $6\times10^{12}$ $Z$ bosons.}\label{llplife}
\end{figure}

From the total width, the decay length of the \dpho~ $L_{\dpho}$ 
can be calculated as:
$$
L_{\dpho}=\frac{\sqrt{\gamma^2-1}}{\Gamma_{\dpho}},
$$
where $\gamma$ is the relativistic boost factor of the \dpho.
The value of $L_{\dpho}$ in the $m_{\dpho}-\epsilon$ plane calculated for
the process $e^+e^-\rightarrow\gamma\dpho$  at $\sqrt{s}=91.2$~GeV
is shown in Figure~\ref{llplife}.

From this plot the theoretical parameter space accessible 
to FCC-ee searches can be evaluated. The blue line shows the values
of $m_{\dpho}$ and $\epsilon$ yielding 3 events for the $Z$-pole run.
Over the full mass range up to the $Z$ mass, in absence of background, the minimum
reachable value of $\epsilon$ is a few $10^{-5}$.

The approximate region over which the  $\dpho$ decays inside the detector with 
an observable flight path is comprised between the red line, corresponding  
to $L_{\dpho}=0.5$~mm, and the green line,  corresponding to $L_{\dpho}=4$~m. 
A useful number of long-lived events may thus be observed inside
the triangle bounded from  above by the red line, and from below by the blue line,
which extends up to $m_{\dpho}$ of a few GeV.
This region has discontinuities for masses where the \dpho~ mainly decays into 
hadronic resonances, due to the combined effect of lower  production rate for 
the muon channel and of shorter lifetime.

\section{Signal and background generation}\label{sec:samples}
%\section{Signal and background generation}
The signal samples were generated with \verb+MG5aMC@NLO+ \cite{Alwall:2014hca}, based on the 
model files made available in \cite{HAHM}. The model includes both a dark photon
and a dark Higgs, and has four parameters, $m_{\dpho}$ and $\epsilon$ for the
dark photon sector, the mass of the dark Higgs $H_s$, and its coupling $\kappa$.
The latter two parameters are set respectively at 1~TeV and $10^{-9}$, decoupling
the dark Higgs sector from the analysis.

The generated process was:
$$
e^+e^-\rightarrow \gamma A^{\prime}, A^{\prime}\rightarrow \mu^+\mu^-
$$
Scans were performed over $m_{\dpho}$ between 0.5 and 360 GeV, 
generating for each of the centre-of-mass energies of 91.2, 160,  240 and 365 GeV 
a grid of the kinematically accessible mass values.
The mixing of the \dpho~ with the photon $\epsilon$ was set to $10^{-4}$, near
the minimum value of $\epsilon$ explored in the analysis. A total of 100k events 
per point were generated for these samples.

For the long-lived analysis, samples of 100k events for  masses between 0.4 and 1.6~GeV
with a granularity of 0.1 GeV were generated, for a grid of values
of $\epsilon$ varying between $10^{-3.5}$ and $10^{-5}$ at $\sqrt{s}=91.2$~GeV.

The parton-level events were hadronised with \verb+PYTHIA8+ \cite{Sjostrand:2014zea} 
and then fed into the \verb+DELPHES+ \cite{deFavereau:2013fsa} fast simulation of the IDEA Detector \cite{IDEAStudyGroup:2025gbt}, based on the official datacards used for the  "Winter2023" production of the FCC-PED study \cite{winter2023setup}.

%For  the backgrounds from $Z$ decays, the official samples produced
%by the central software group for the FCC under the tag
%"Winter2023" were used \cite{winter2023samples}.

For the prompt analysis the irreducible background from the process 
$$
e^+e^-\rightarrow \gamma \mu^+ \mu^- 
$$
was produced at leading order (LO) with \verb+MG5aMC@NLO+.
The generation-level  requirements were 
that photons and leptons be produced within a pseudorapidity $\eta$ of $\pm2.6$,
corresponding to a polar angle of approximately $10^{\circ}$ from the beam,
and that the photon (muon) energy be in excess of 5 (0.5)~GeV, respectively.
Samples of 10M events were produced for each of the considered centre of mass energies,
91.2, 160, 240 and 365~GeV. For the $Z$-pole run an additional 10M background sample
with the requirement $m_{\mu^+\mu^-}<15$~GeV was generated to ensure enough 
Monte Carlo statistics in the $m_{\mu^+\mu^-}$ mass bins below 20~GeV. 

For the long-lived analysis, a sample of 50M $e^+e^-\rightarrow \gamma \mu^+ \mu^-$ 
events were generated at 91.2~GeV, with the additional requirement that 
$m_{\mu^+\mu^-}<4$~GeV, and $44.5<E_\gamma<46.5$~GeV. The cross-section for this
sample is 0.63~pb, and the generated statistics corresponds to an integrated luminosity 
of 80~ab$^{-1}$.

Two samples of 10M events for the processes $e^+e^-\rightarrow \gamma \tau^+ \tau^-$
and $e^+e^-\rightarrow \gamma b \bar{b}$ with the same requirements on the
photon energy were also generated to check the reducible backgrounds to the long-lived
analysis. For both samples the generated statistics corresponds to several times
the expected statistics for the $Z$-pole run.

All the background samples were then processed through 
the same \verb+PYTHIA8-DELPHES+ chain as the signal events.

\section{Detector simulation}\label{sec:detector}
%The IDEA detector concept for the LHC is the proposal for a general-purpose 
%detector for the FCC-ee. 
The IDEA detector concept is a proposal for a general-purpose detector for the FCC-ee.
The design  includes
an inner detector composed of 5 Monolithic silicon pixel (MAPS) layers
followed by a high-transparency and  high-resolution drift chamber.  
Outside the inner detector is located a high-resolution dual readout 
crystal calorimeter, surrounded by a superconducting solenoid producing a 2~T magnetic
field.
Outside the solenoid is  a  high granularity fiber dual-readout hadronic calorimeter, 
followed by three $\mu$-rwell  layers for muon detection, embedded in the 
return yoke of the solenoid.

The present study relies on the parameterised simulation in \verb+DELPHES+ of the 
inner detector for the estimate of the tracking and vertexing performance
of IDEA, and on a simple energy response parametrisation for the electromagnetic
calorimeter.

The vertex detector is simulated as 5 cylindrical layers with 
radius between 1.2 and 31.5 cm with 2-D readout, and resolution $3~\mu$m
except the outermost layer with a resolution of $7~\mu$m,
and 3 disks in each of the forward/backward regions with a resolution of $7~\mu$m.
The drift chamber is modelled as 112 co-axial layers, arranged in 24 identical azimuthal sectors, at alternating-sign stereo angles ranging from 50 to 250 mrad, and
an assumed resolution on the single measurement of $100~\mu$m. The chamber has
a length of 4 meters, and covers the radius between 34~cm and 2~m, yielding 
full tracking efficiency for 
polar angles $\theta$ larger than about $10$ degrees.
The DELPHES simulation software relies on a full description of this geometry and accounts for the finite detector resolution and for the multiple scattering in each tracker layer. Based on the measured `hits' in the different layers, it turns charged particles emitted within the angular acceptance of the tracker into five-parameter tracks (the helix parameters that describe the trajectory of the particle, including the transverse and longitudinal impact parameters), and determines the full covariance matrix of these parameters. Details on the procedure are provided in
\cite{Bedeschi:2022rnj}.
Vertices are reconstructed using these tracks as input, based on a simple $\chi^2$ minimisation with constraints, yielding 3D
vertexes with their $\chi^2$ ($\chi^2_{vx}/ndf$) and covariance matrix. More details on the vertexing code used here can be found in~\cite{Bedeschi:2024uaf}.

For the crystal electromagnetic calorimeter a gaussian smearing is applied to deposited electromagnetic energy with the resolution parametrised as \cite{Lucchini:2022vss} ($E$ in GeV):
\begin{equation}
\frac{\sigma(E)}{E}=\frac{0.03}{\sqrt{E}}\,\oplus\,0.005\,\oplus\,\frac{0.002}{E}\label{eq:calres}. 
\end{equation}

\section{Analysis}\label{sec::ana}
The final state for the signal addressed in the present analysis
$$
e^+e^-\rightarrow \gamma A^{\prime}, A^{\prime}\rightarrow \mu^+\mu^-
$$
includes two opposite charge muons and a photon. The invariant mass of the
two muons $m_{\mu^+\mu^-}$ has a narrow peak at the mass of the \dpho~,
with the width of the peak dominated by the experimental resolution
on the reconstruction of $m_{\mu^+\mu^-}$. 

The photon recoiling against a dark photon of mass $m_{\dpho}$ has a fixed 
energy $E_{rec}$, determined by the recoil formula:
\begin{equation}\label{eq:recoil}
	E_{rec}=\frac{E_{CM}^2-m_{\dpho}^2}{2\,E_{CM}},
\end{equation}
where $E_{CM}$ is the center-of-mass-energy in of the collisions.

The starting requirement in the analysis is the presence of exactly two 
reconstructed muons with energy above 2 GeV and of a photon with 
energy in excess of 5 GeV within the angular acceptance of the
tracking detector.

In order to suppress reducible backgrounds, a preselection is applied
requiring that the invariant mass of the two muons and the photon
be larger than 90~GeV, and that the two muon tracks form a good 
vertex ($\chi^2_{vx}/ndf<10$)  inside the geometrical acceptance  of the 
IDEA inner detector.

Two different analyses are developed from this point, with different 
strategies for the separation of signal and background, a prompt 
search, and a long-lived search. 

The prompt search applies no requirement on the distance of the 
reconstructed $\mu^+\mu^-$ vertex from the interaction point $d_{vx}$ , and 
relies on a detailed analysis of the final state kinematics to 
suppress the large $e^+e^-\rightarrow \gamma \mu^+ \mu^-$ SM background. 
This analysis covers the complete mass range from 0.4 GeV to 360~GeV.

The long-lived search addresses the case where the \dpho~ decays 
after an observable path in the detector. A selection on $d_{vx}$
can in principle fully eliminate the dominating SM prompt background, leaving 
only decays from heavy flavours which have a very different final state
kinematics from the signal.  As shown in Figure~\ref{llplife},
this branch of the analysis only covers a limited mass range, 
but may probe lower values of the couplings than the prompt analysis.

In the following sections the two analyses will be described in detail.

\subsection{Prompt analysis}\label{sec::prompt}
For the prompt analysis, after the preselection is applied, 
the background is expected to be dominated by the SM process
$e^+e^-\rightarrow \gamma \mu^+ \mu^-$ which is abundantly produced,
and has the same final state as the signal.

The initial step in the selection is the requirement, 
for each dark photon test mass $m_{\dpho}$ considered,
that the event kinematics is compatible with a photon recoiling 
against a massive particle with mass $m_{\dpho}$.

\begin{figure}
\centering
\includegraphics[width=0.6\textwidth]{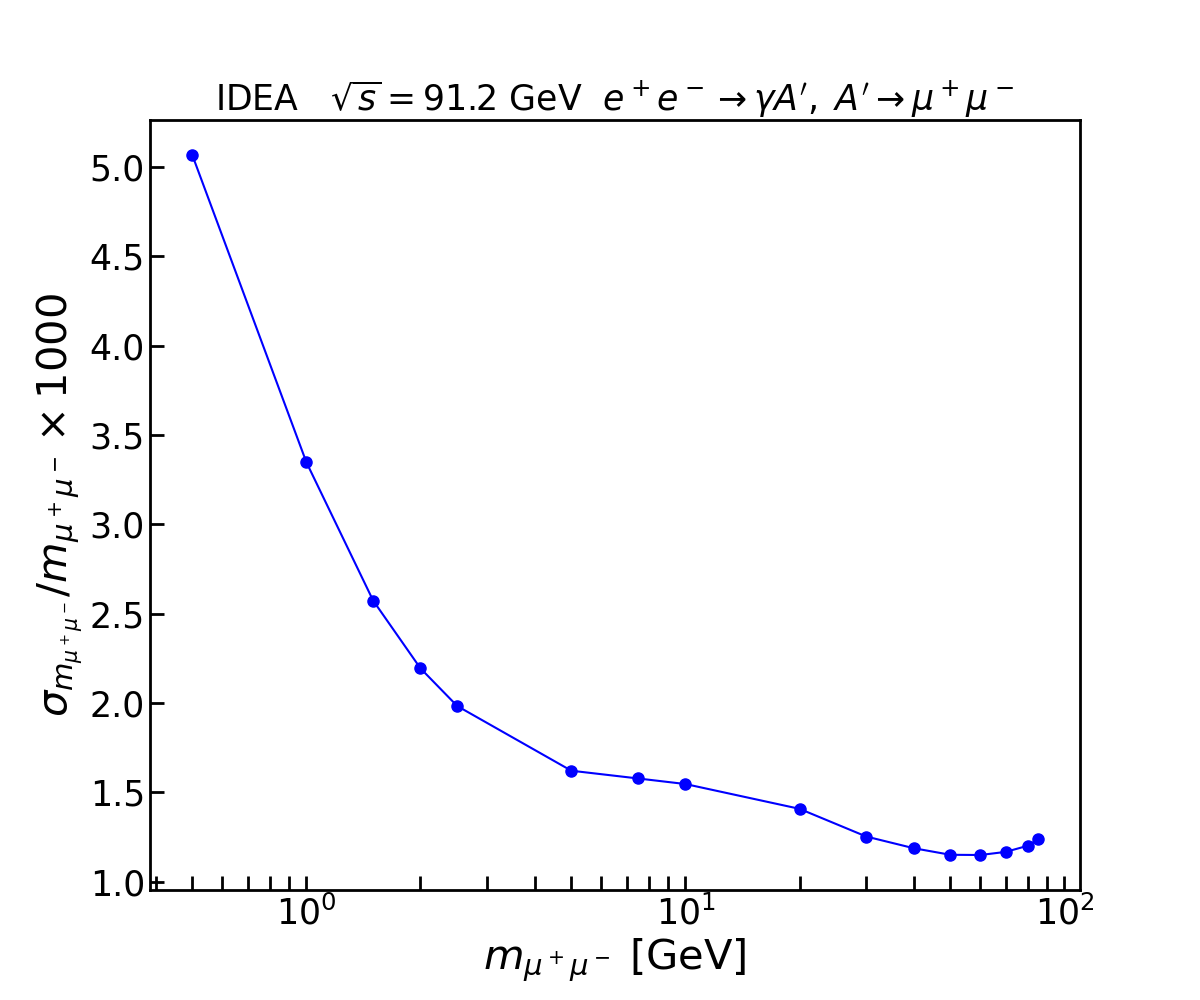}
	\caption{Relative mass resolution for muon pairs from the of \dpho~  multiplied by 1000 as a function of $m_{\mu^+\mu^-}$.}\label{fig:sigmamu}
\end{figure}

To this effect, for each event the  variable $M_{cut}$ can be built as:
\begin{equation}
	M^2_{cut}=\frac{(m_{\mu^+\mu^-}-m_{\dpho})^2}{\sigma(m_{\mu^+\mu^-})^2}+\frac{(E_{\gamma}-E_{rec})^2}{\sigma(E_{\gamma})^2},
\end{equation}
where $E_{rec}$ is defined in Equation~\ref{eq:recoil}, and $\sigma(m_{\mu^+\mu^-})$ and $\sigma(E_{\gamma})$ are respectively the expected
muon-muon mass resolution, and photon energy resolution. The dependence  of 
$\sigma(E_{\gamma})$ on $m_{\mu^+\mu^-}$ can be calculated as  the convolution of  
Eq.~\ref{eq:calres} and Eq.~\ref{eq:recoil}.
It is approximately flat at  0.7\% up to $m_{\mu^+\mu^-}\sim 40$~GeV and shoots up to
a few percent as $m_{\mu^+\mu^-}$ approaches $m_{Z}$.
The value of $\sigma(m_{\dpho})$ depends on the kinematics of the events. Its dependence on $m_{\mu^+\mu^-}$ is shown in Figure~\ref{fig:sigmamu}.
The resolution varies smoothly between 0.1 and 0.15\% above 5~GeV, and increases up to 0.5\% at 0.5~GeV, when the uncertainty on the distance between the two muon tracks becomes the dominant factor in the mass measurement. 

\begin{figure}
\centering
\includegraphics[width=0.45\textwidth]{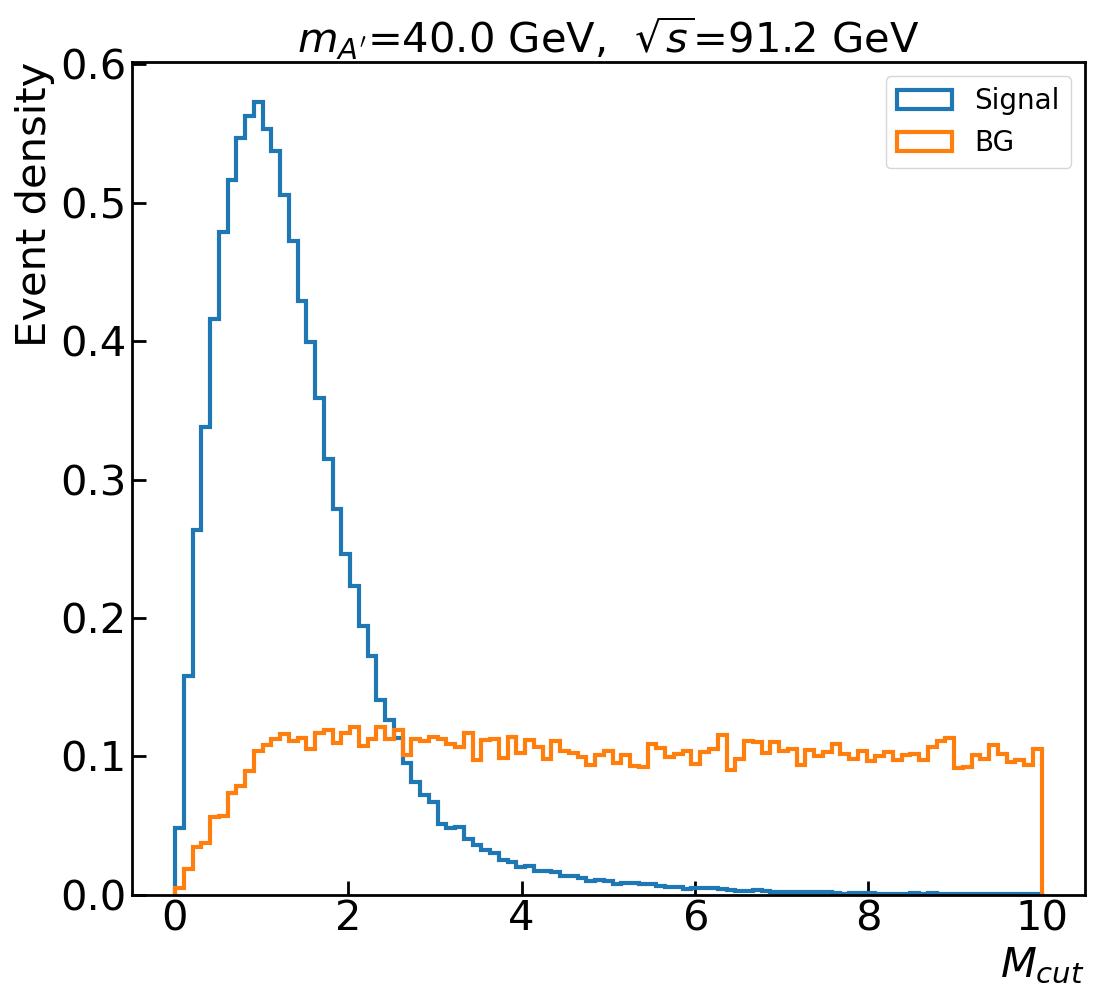}
\includegraphics[width=0.45\textwidth]{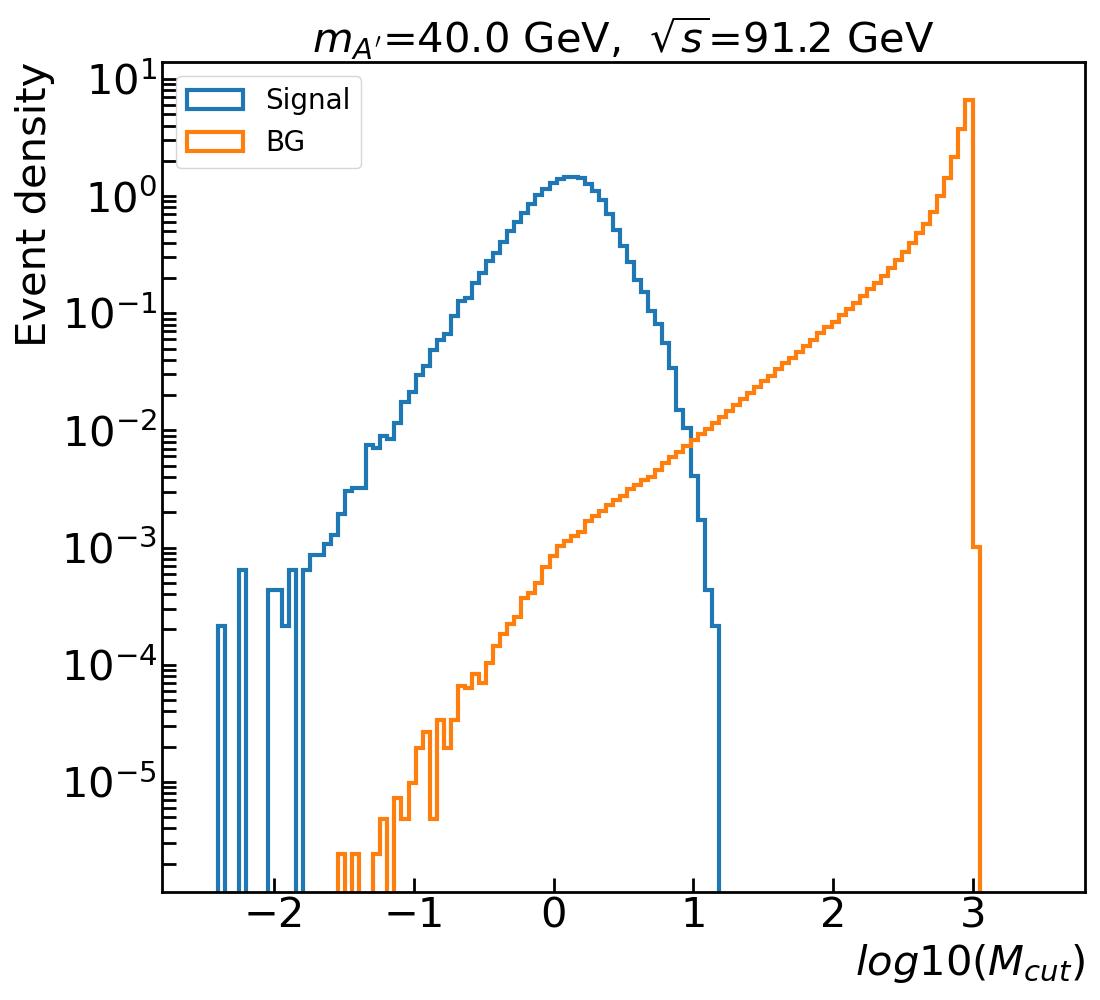}
	\caption{Distribution of the variables $M_{cut}$ truncated at 10 (left) 
	and $\log_{10}(M_{cut})$ for
	signal (blue) and background (orange) for $m_{\dpho}=40$ GeV.}\label{fig:sigmam}
\end{figure}

The discriminant variable $M_{cut}$ is approximately
independent of $m_{\dpho}$, and has an excellent signal to background separation power.
The distributions of $M_{cut}$, truncated at a value of 10, and of $\log_{10}(M_{cut})$
are shown for $m_{\dpho}=40$~Gev in Figure~\ref{fig:sigmam}, for both signal and 
background at $\sqrt{s}=91.2$~GeV. The selection $M_{cut}>2$ approximately optimises the signal/background separation at all masses.

%A request of $\sigma(m_{test})<2.5-3$ 
%keeps most of the signal and rejects a large fraction of the background.
Given the relative values of the two components, and their
dependences on $m_{\mu^+\mu^-}$, the rejection power of $M_{cut}$  
will be dominated by $\sigma(m_{\mu^+\mu^-})$ except for the very lowest values 
of $m_{\dpho}$. The sensitivity of the analysis is therefore mainly determined by the
performance of the tracking system.

Besides the selection on $M_{cut}$, additional separation power
between signal and background can be obtained through a detailed study
of the angular distributions of the final state particles.

\begin{figure}
\centering
\includegraphics[width=0.45\textwidth]{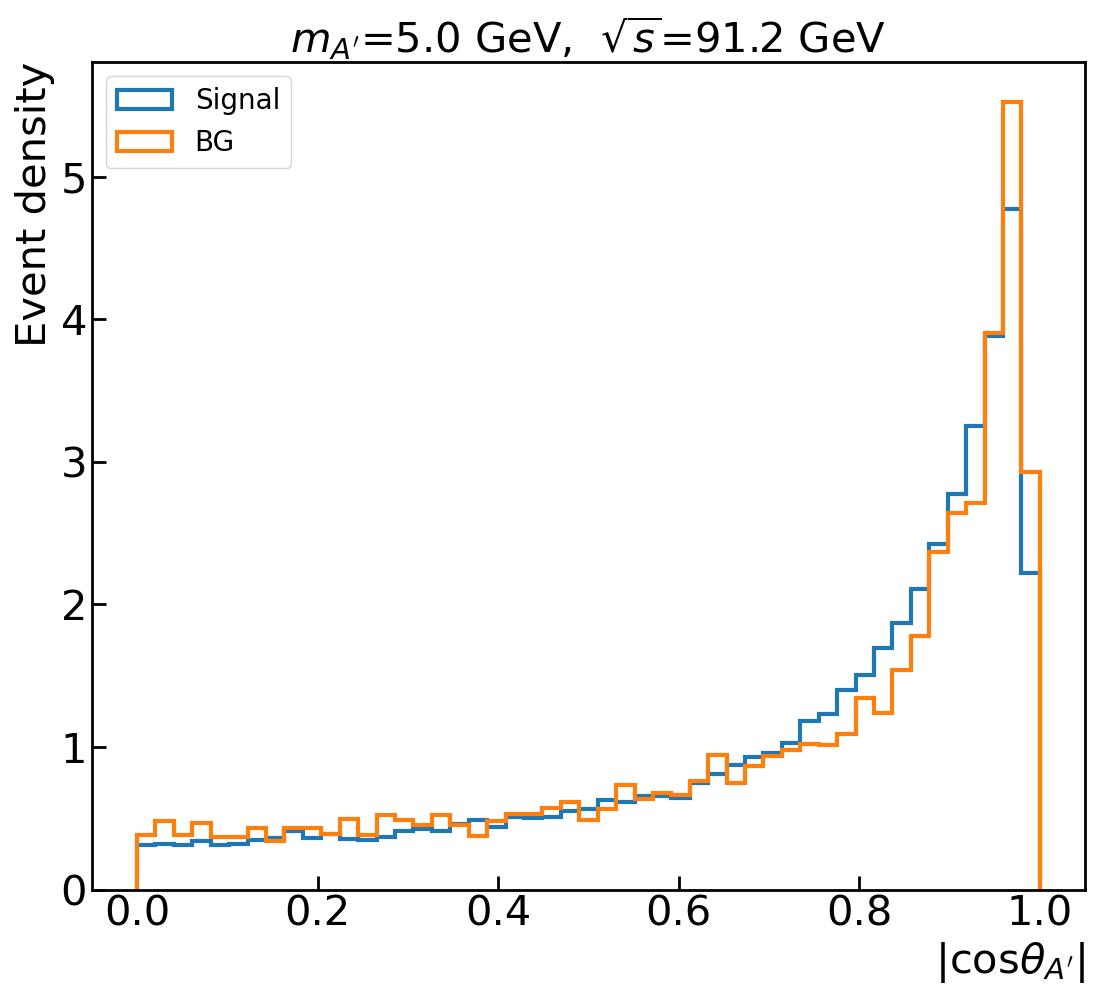}
\includegraphics[width=0.45\textwidth]{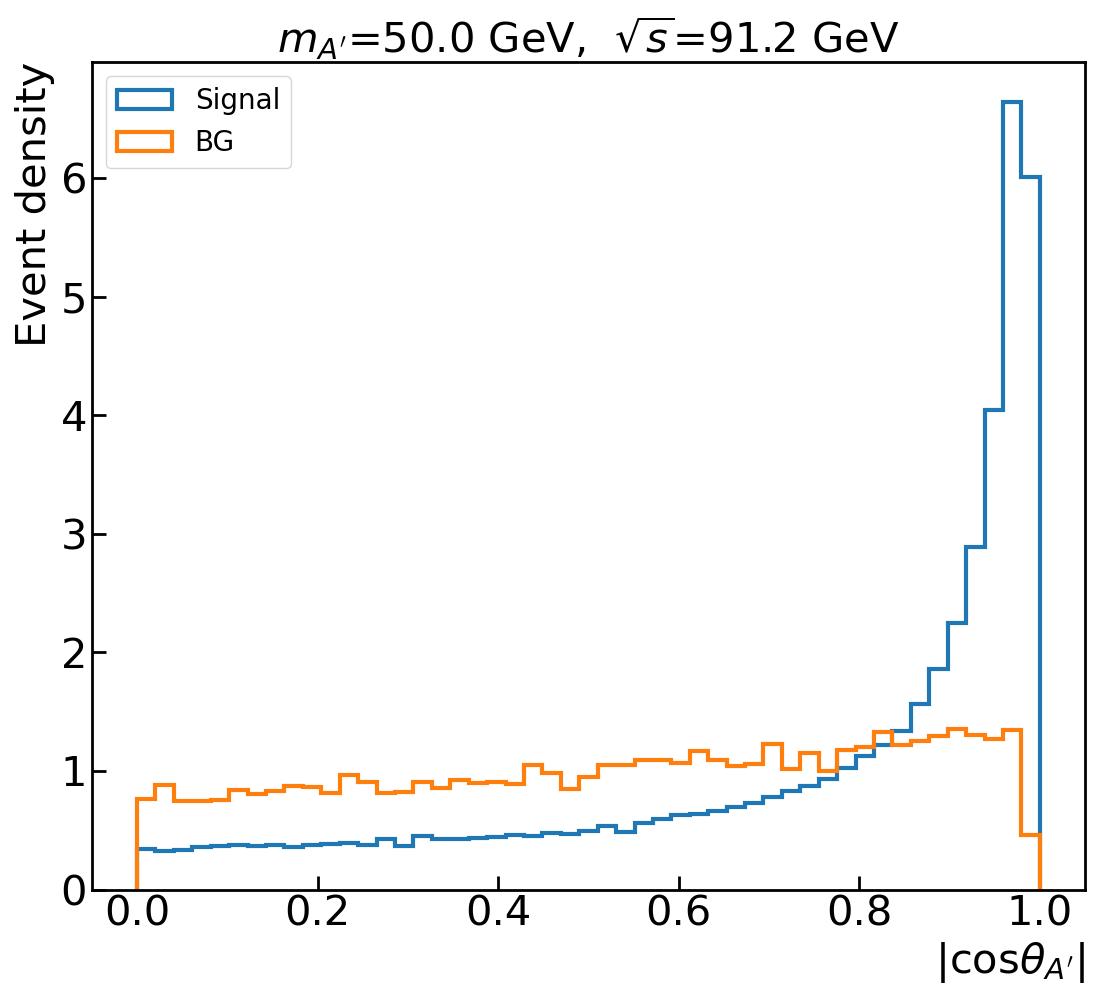}
\caption{Distribution of the variable $|\cos\theta_{\dpho}|$ for
	signal (blue) and background (orange) for two values of $m_{\dpho}$, 5 (left) and 50 (right) GeV.} \label{fig:cthdpho}
\end{figure}

\begin{figure}
\centering
\includegraphics[width=0.45\textwidth]{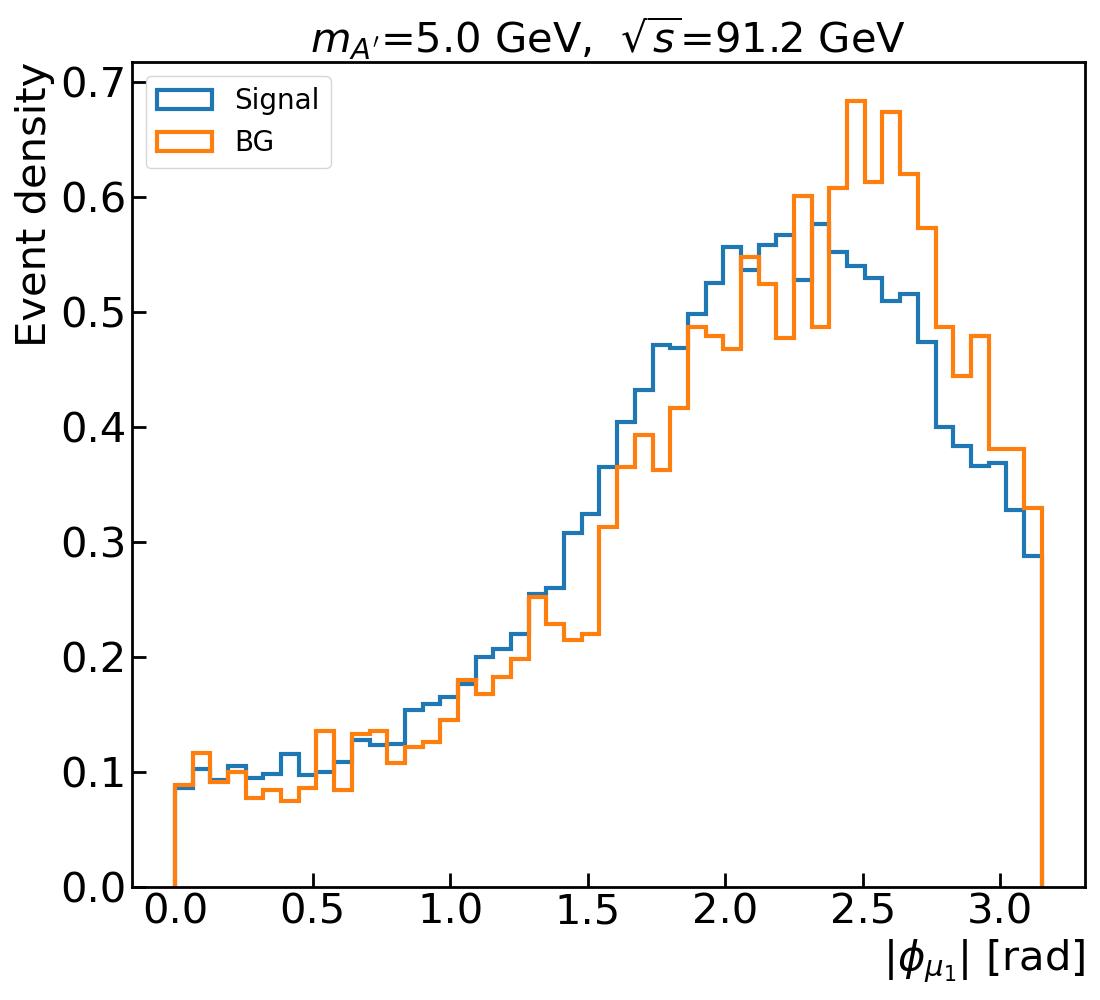}
\includegraphics[width=0.45\textwidth]{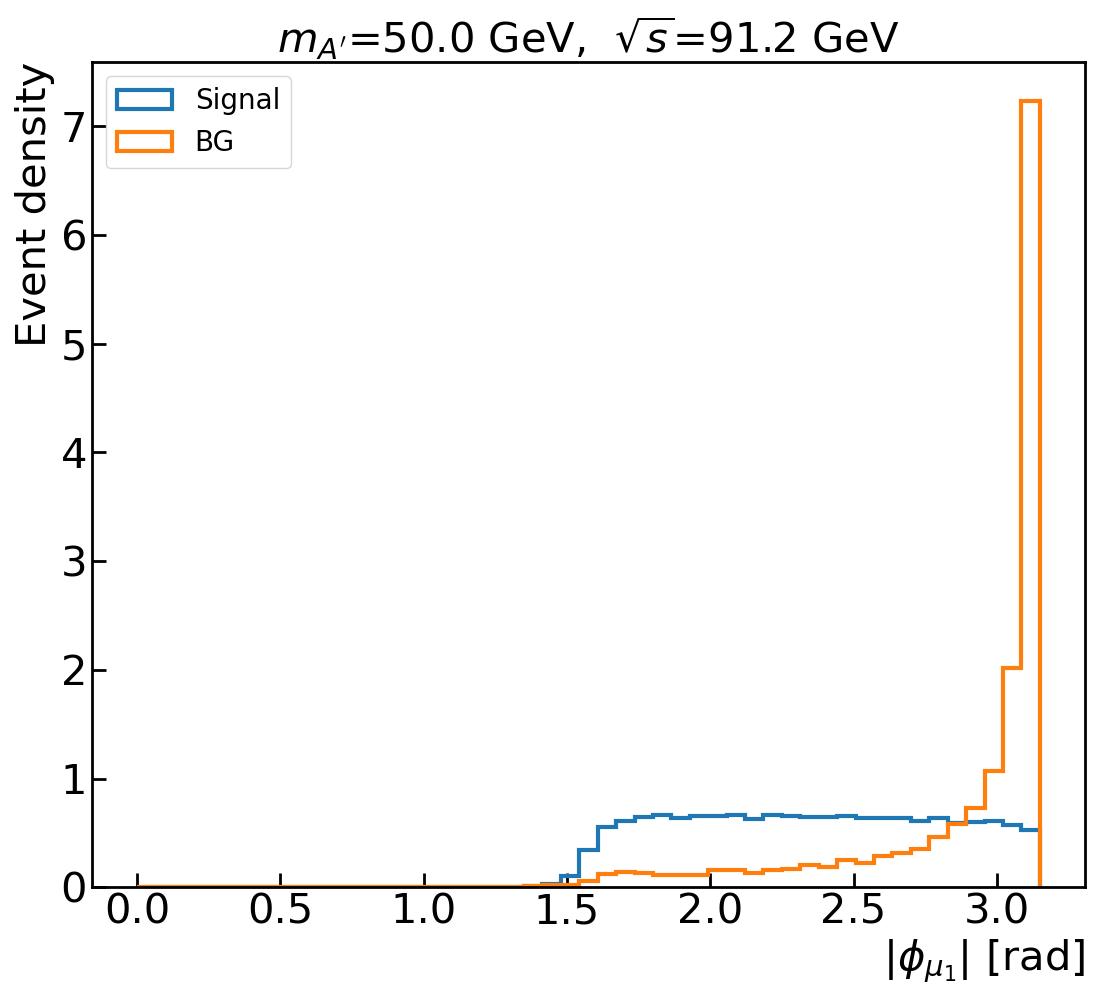}
\caption{Distribution of the variable $|\phi_{\mu_1}|$  calculated 
	in the rest frame of the ALP for
        signal (blue) and background (orange) for two values of $m_{\dpho}$, 5 (left) and 50 (right) GeV.} \label{fig:phim1}
\end{figure}

\begin{figure}
\centering
\includegraphics[width=0.45\textwidth]{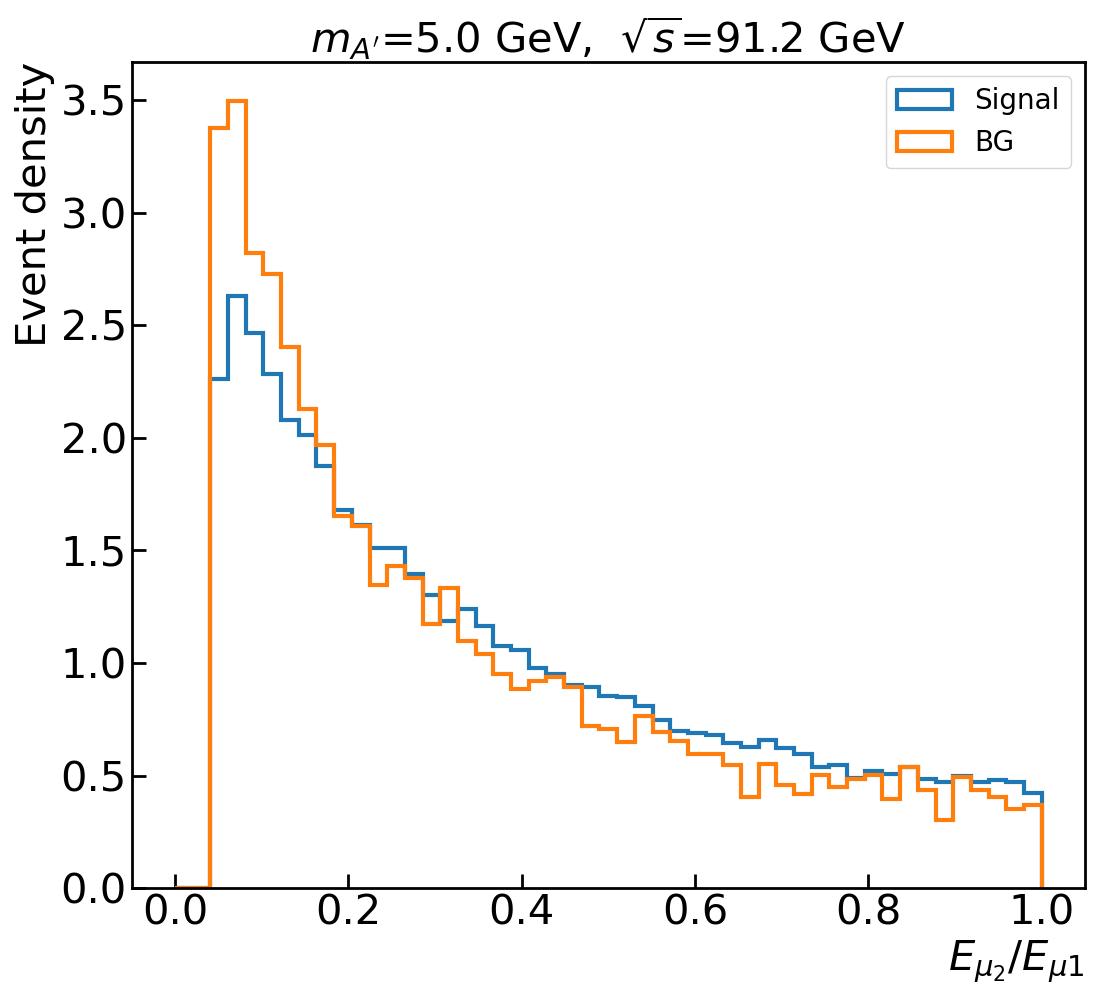}
\includegraphics[width=0.45\textwidth]{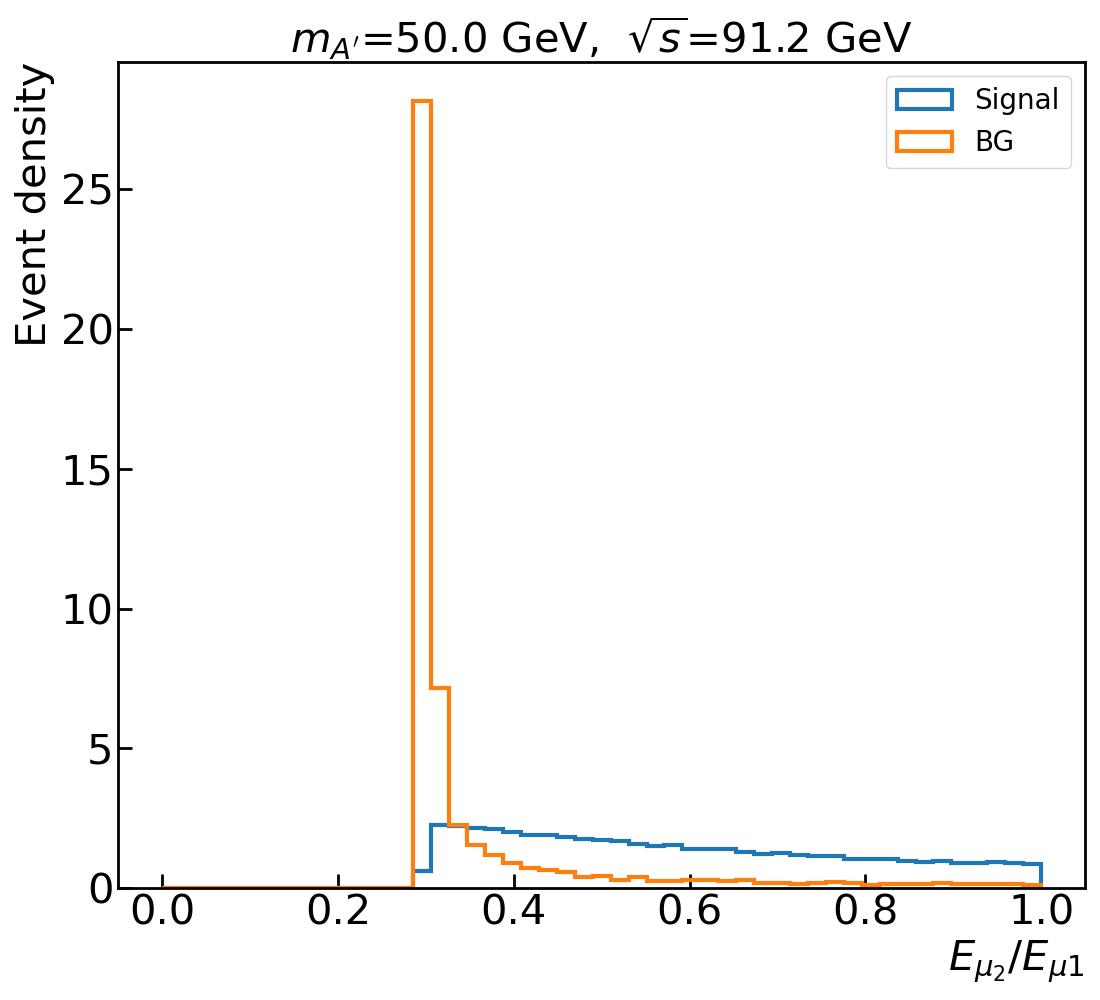}
\caption{Distribution of the variable $E_{\mu_2}/E_{\mu_1}$  calculated for
        signal (blue) and background (orange) for two values of $m_{\dpho}$, 5 (left) and 50 (right) GeV.} \label{fig:emr12}
\end{figure}

\begin{figure}
\centering
\includegraphics[width=0.45\textwidth]{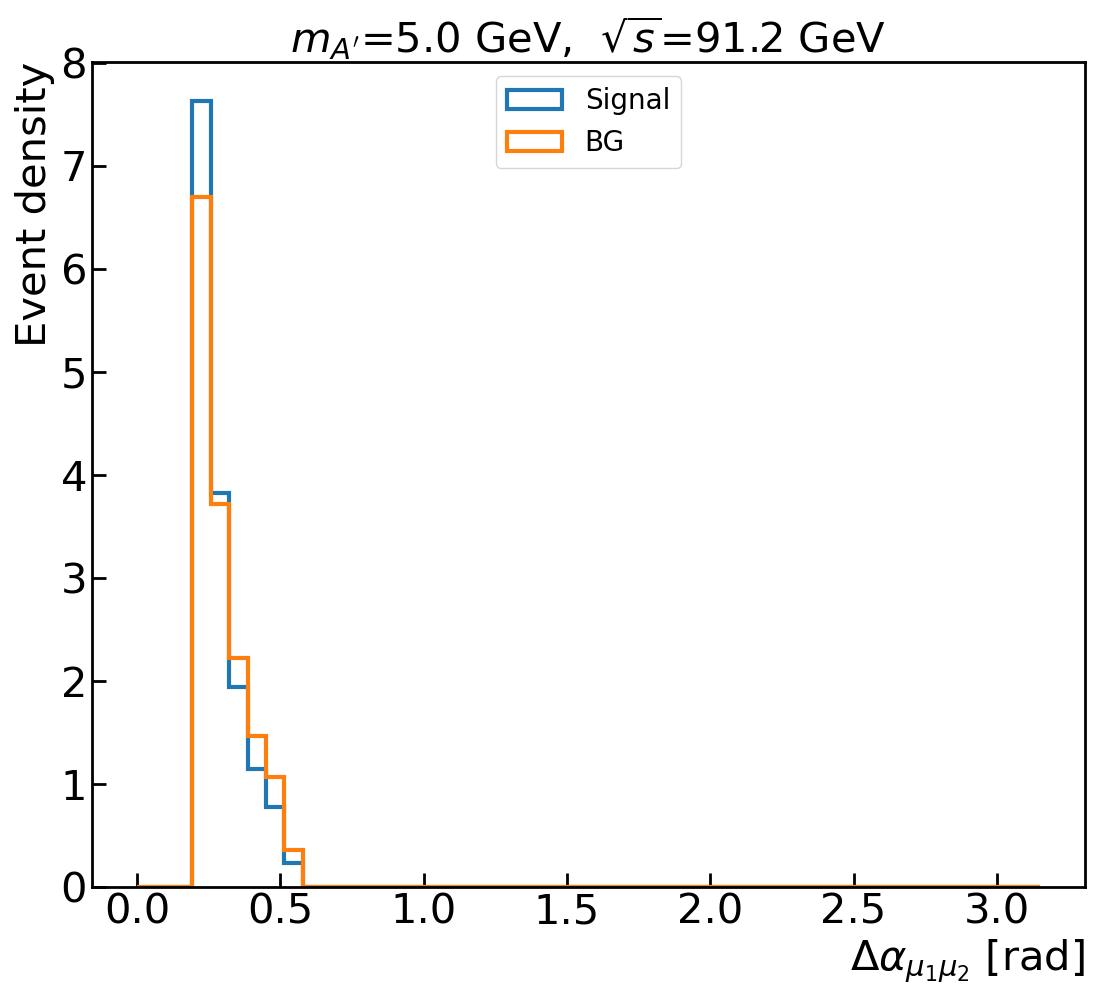}
\includegraphics[width=0.45\textwidth]{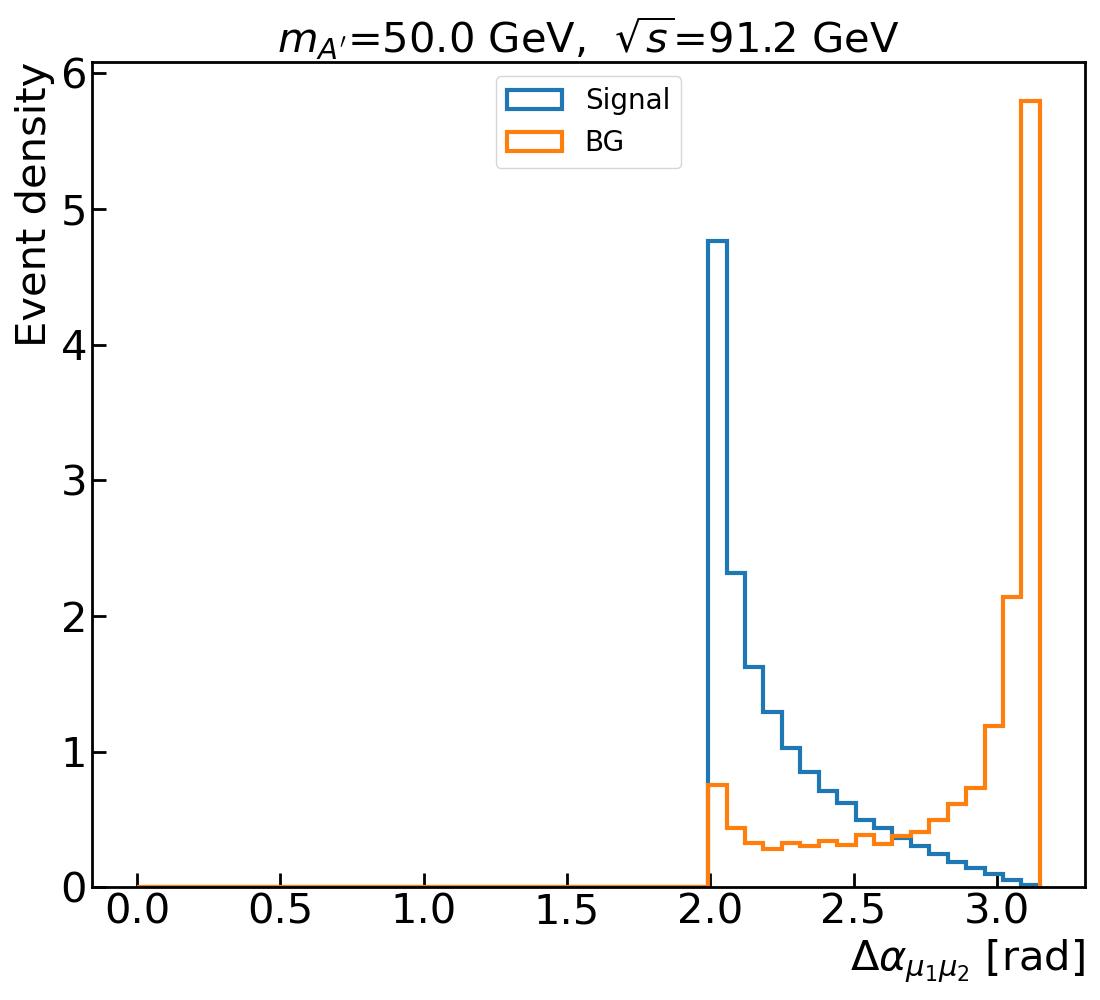}
\caption{Distribution of the variable $\cos\Delta\alpha_{\mu_1\mu_2}$  calculated for
        signal (blue) and background (orange) for two values of $m_{\dpho}$, 5 (left) and 50 (right) GeV.} \label{fig:dang12}
\end{figure}

Each event is fully defined by 9 variables, i.e. the three-momenta of the
two muons and the photon. The energy and momentum constraints from the $e^+e^-$ collision reduce the independent variables to five.
The signal is a photon accompanying a particle undergoing a two-body decay, and a convenient choice
of variables is the invariant mass of the $\mu^+\mu^-$ pair, the angular coordinates $\theta_{\gamma}$ and $\phi_{\gamma}$ of the 
photon from the $e^+e^-\rightarrow\gamma\dpho$ production in the lab frame, and  $\theta_{\mu_1}$ and $\phi_{\mu_1}$, the  coordinates of the leading muon, calculated in the rest frame of the $\mu^+\mu^-$ pair. Given the cylindrical symmetry of the system around the beam axis, the events can be rotated in such a way that $\phi_{\gamma}=0$.  After this rotation, the event is fully defined by $m_{\mu^+\mu^-}$ and by three angular variables. A reasonable choice is $\cos\,\theta_{\gamma}$=$-\cos\,\theta_{\dpho}$, $\cos\,\theta_{\mu_1}$ and $\phi_{\mu_1}$. The latter two are calculated in the  \dpho~ rest frame. After inspecting the variables, $\cos\,\theta_{\mu_1}$ was replaced by two variables in the laboratory frame that encode equivalent information on the  \dpho~ decay kinematics: the ratio of the energies of the two muons, $E_{\mu_2}/E_{\mu_1}$, and the cosine of their angular distance in 3-d space, $\cos\Delta\alpha_{\mu_1\mu_2}$, which provide better background discrimination.
The distributions of the four discriminant variables defined above are shown in Figures~\ref{fig:cthdpho}-\ref{fig:dang12} for two different test masses and for both  signal and background for $\sqrt{s}=91.2$~GeV. The selection $M_{cut}<2$ has been
applied for both signal and background.

The discrimination power is higher for $m_{\dpho}=50$~GeV than for $m_{\dpho}=5$~GeV. The difference is due to the 
fact that for masses much lower than $m_{Z}$  the $\mu^+\mu^-$ pair results from the decay 
of a particle with photon-like couplings \cite{Curtin:2014cca}, and the
background is dominated by a virtual photon going into $\mu^+\mu^-$.
When $m_{\dpho}$ becomes comparable to $m_{Z}$, there is an admixture of photon and 
$Z$ contributions with a relative ratio which evolves in a different way for 
signal and background, generating a difference in the angular distributions.
For similar reasons, for values of $\sqrt{s}$ different from the $Z$-pole, the analysis based on angular variables is in general less powerful than in the
case of the $Z$-pole run.
%The shape of the distributions display a mild dependence from the test mass,
%differently from lab-frame variables like the angle between $\gamma_1$ and $\gamma_2$,
%or the ratio of their energies.

\begin{figure}
\centering
\includegraphics[width=0.495\textwidth]{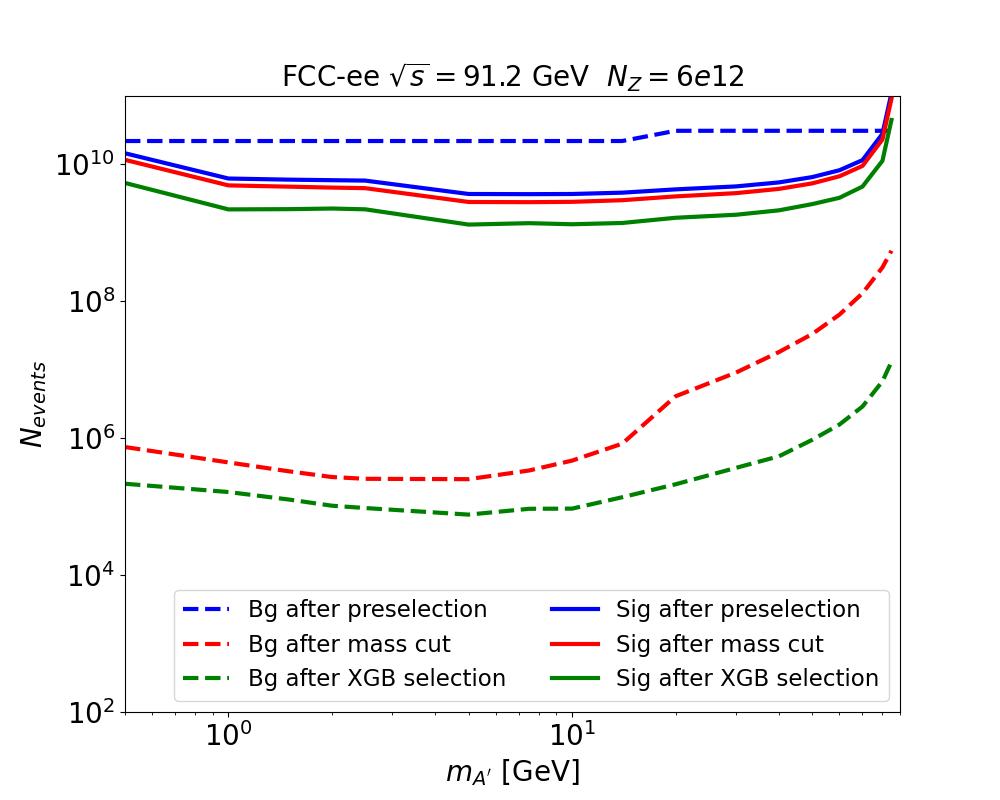}
\includegraphics[width=0.495\textwidth]{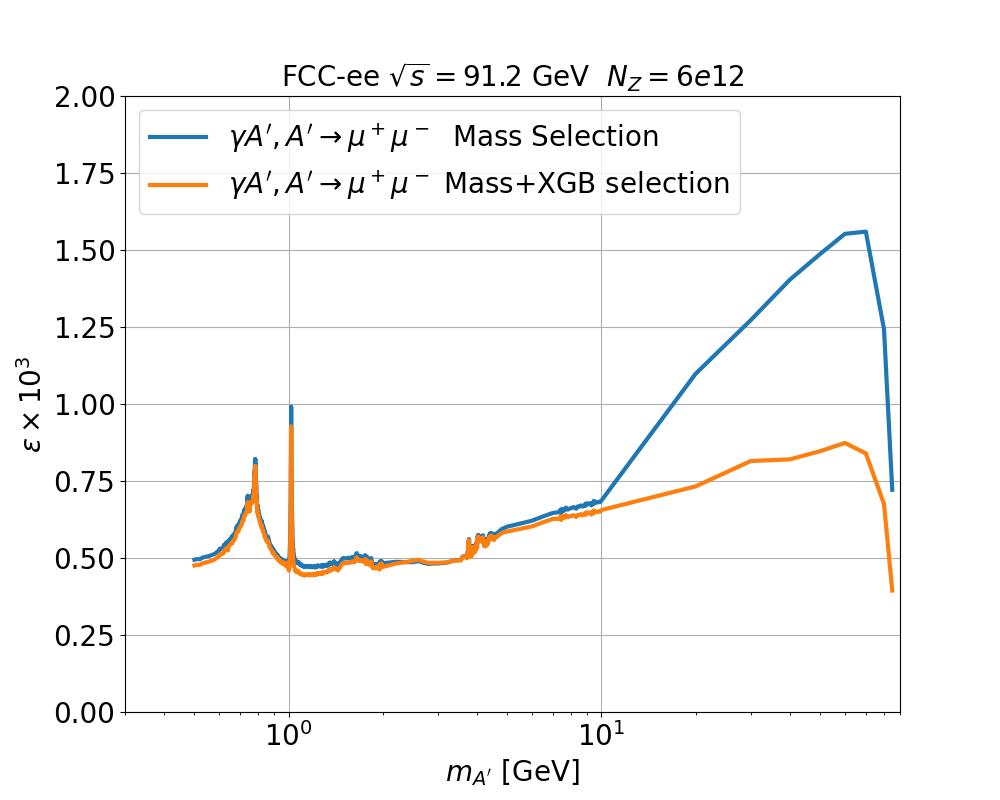}
\caption{Left: number of expected signal (full line) and background (dashed line) events for the FCC-ee $Z$-pole 
run and each of the selection steps described in the text. The lines after the XGB selection are given for a selection on $P_{XGB}$ yielding an efficiency for the signal of 50\% with respect to the mass selection. Right: experimental 95\% reach on the mixing parameter $\epsilon$ as a function of $m_{\dpho}$. The blue line gives the result after the mass cut, and the orange
line after the XGB cut.
}\label{fig:nevbg}
\end{figure}

For each test mass a signal selection is developed, starting from the preselection described above, based on the following steps:
\begin{itemize}
\item
A boosted decision tree (XGBoost \cite{xgboost}, XGB in the following) is
trained on four variables, $|\cos\Delta\alpha_{\mu_1\mu_2}|$, $E_{\mu_2}/E_{\mu_1}$,
$|\cos\theta_{\dpho}|$, and $|\phi_{\mu_1}|$, where the last variable is in the rest frame
of the ($\mu_1,\mu_2$) system.
The distributions of the discriminant variables depend on $m_{\dpho}$. Therefore, for each test value of $m_{\dpho}$, the samples of signal and background events used for training are requested to satisfy the condition $|m_{\mu_1\mu_2}-m_{\dpho}|<5$~GeV ($M_{cut}<10$)
for $m_{\dpho}>(\le)10$~GeV respectively. The selections are aimed
at ensuring a final state kinematics as close as possible to the one for the target
signal, while retaining enough Monte Carlo events in the training and test 
samples as to ensure a statistically stable selection.
\item
The XGB probability ($P_{XGB}$)  is then calculated on signal and
background test samples for each target $m_{\dpho}$. 
An example distribution of the  XGB probability ($P_{XGB}$) for $m_{\dpho}=5$ and 50~GeV
is shown in Figure~\ref{fig:XGB} for events with $M_{cut}>2$.
\begin{figure}
\centering
\includegraphics[width=0.45\textwidth]{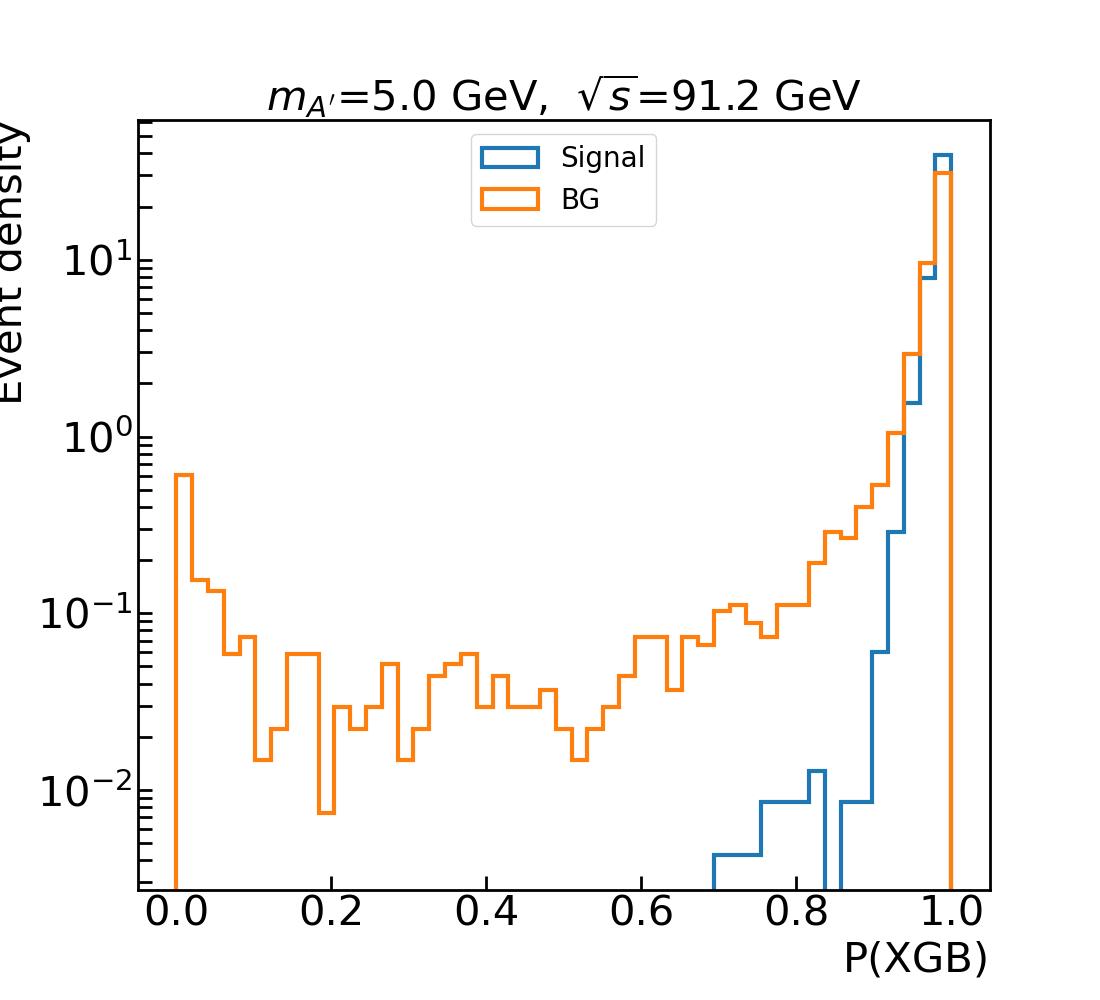}
\includegraphics[width=0.45\textwidth]{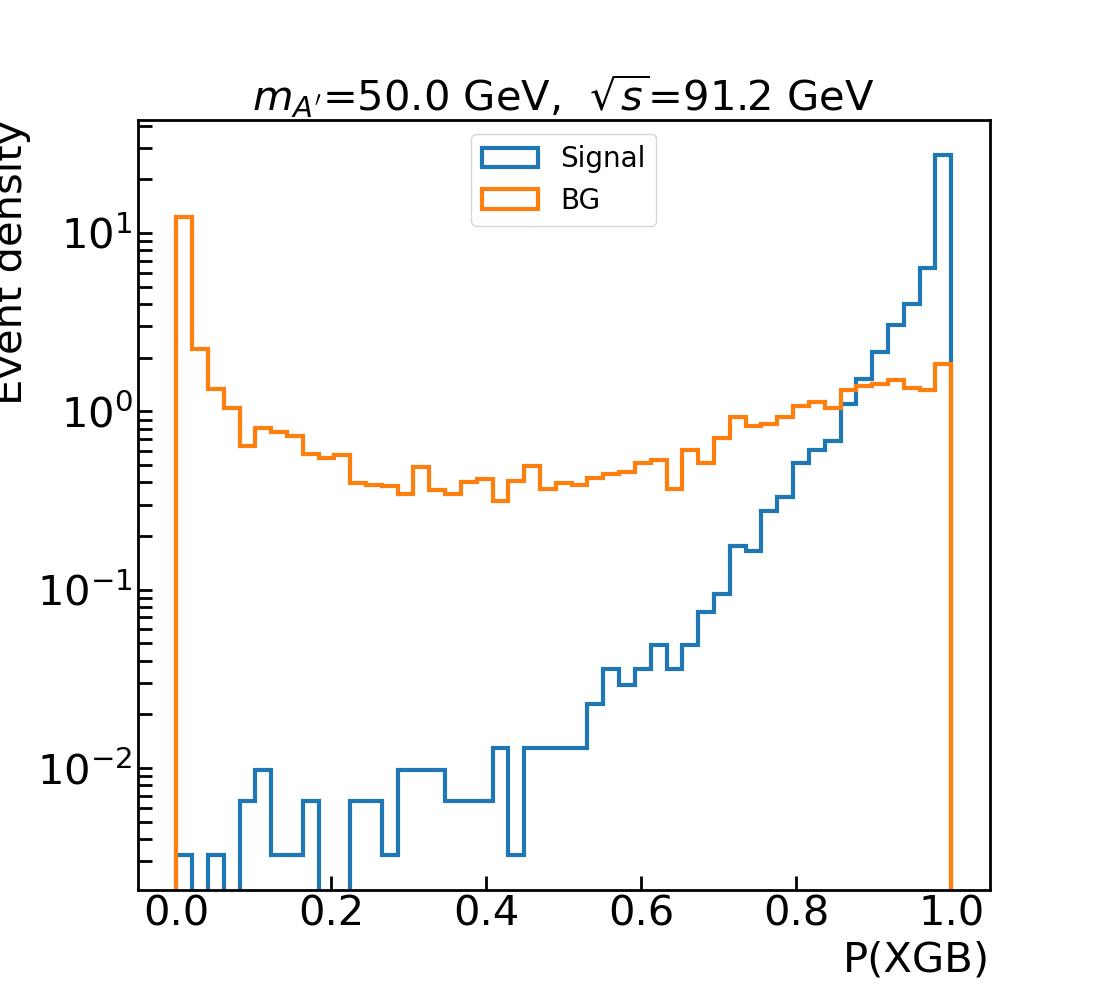}
	\caption{Distribution of the variable $P_{XGB}$ for
	signal (blue) and background (orange) for $m_{\dpho}=5$(left) and 50 (right) GeV.}\label{fig:XGB}
\end{figure}
\item
For each FCC-ee run, the number of events in the test sample 
is normalised to the expected running statistics. For $m_{\dpho}<10$~GeV 
the value of $BR(\dpho\rightarrow\mu^+\mu^-)$ calculated in \cite{Curtin:2014cca}
		is used.  The statistical significance $Z_{stat}$, defined as in \cite{ATLASstat} is 
calculated as a function of $P_{XGB}$ for events passing the mass selection
 $M_{cut}<2$.
\item
For each \dpho~ test mass and each centre-of-mass energy of the FCC-ee,
the cut on $P_{XGB}$ yielding the optimal sensitivity is determined, and
		the value of the coupling $\epsilon$ yielding a statistical significance $Z_{stat}=2$
is calculated.
\end{itemize}
After the XGB selection, the total efficiency for the signal  is $\sim 30-40\%$
over the full considered mass range, whereas the efficiency for the background has a 
strong dependence on $m_{\dpho}$.
The number of expected signal and background events for the FCC-ee $Z$-pole 
run after each of the selection steps described above
is shown in the left side of Figure~\ref{fig:nevbg}. The assumed value of the mixing
for the signal is $\epsilon=1$. The lines after the XGB selection are given for a selection 
on $P_{XGB}$ yielding an efficiency for the signal of 50\% with respect to the previous selection, approximately corresponding to the optimal selection for $m_{\dpho}>20$~GeV.
A strong dependence of the efficiency of the mass cut on $m_{\dpho}$ is observed for the background. For $m_{\dpho}\leq 10$~GeV the background rejection power 
of the XGB selection decreases with decreasing $m_{\dpho}$, and it does not improve the experimental reach below $m_{\dpho}\simeq5$~GeV.
The experimental 95\% CL reach on the mixing parameter $\epsilon$ as a function of $m_{\dpho}$ is
shown on the right side of Figure~\ref{fig:nevbg} for the FCC-ee $Z$-pole run. The result is shown for the mass selection alone and for the mass selection supplemented by the XGB selection, where
the cut on $P_{XGB}$ is set for each test value of $m_{\dpho}$  at the value optimising the sensitivity. A 95\% CL sensitivity limit on $\epsilon$ varying between
$0.5\times10^{-3}$ and $0.85\times10^{-3}$ is achieved throughout the mass range.

\subsection{Long-lived analysis}\label{sec::LLP}
\begin{figure}[h] \centering
\includegraphics[width=0.7\textwidth]{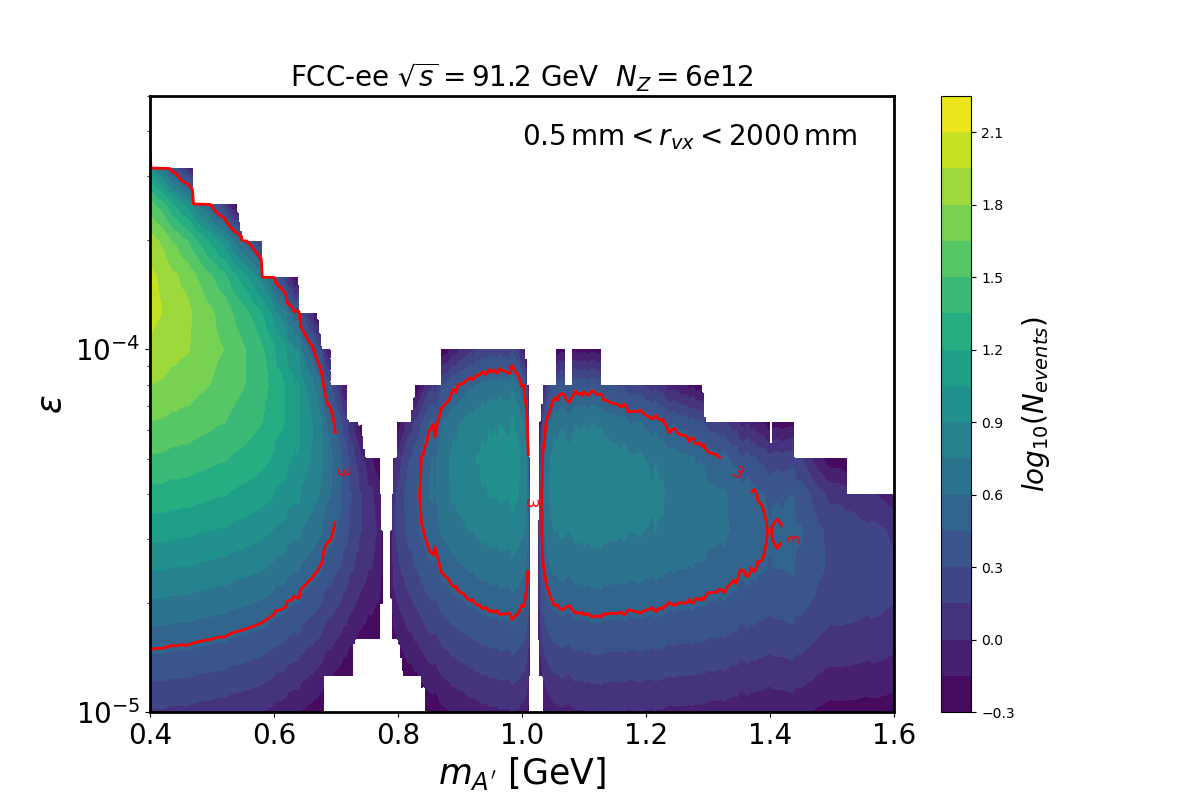}
\caption{Number of events decaying
in the tracker of IDEA at a minimum distance of 0.5~mm from the center of
        the detector for the $Z$-pole run of FCC-ee in the $m_{\dpho}-\epsilon$
        plane. The contour for three events is shown as a red line.}\label{fig:nevllp}
\end{figure}

In Section \ref{sec:model} it was shown that there is a region 
in the $m_{\dpho}-\epsilon$ parameter space where a detectable number of long-lived
dark photons may be produced at the $Z$-pole run of the FCC-ee. 
For the present analysis, it is required that the $\dpho$ decays 
inside the inner the tracker of the IDEA experiment. A quantitative
idea of the accessible area in parameter space can thus be obtained
by calculating for each point  of the $m_{\dpho}-\epsilon$ space the 
integral of the expected number of events with a path in the detector $d_{vx}$ in the 
range 0.5-2000~mm for the expected statistics of the FCC-ee $Z$-pole run.
The results are shown in Figure~\ref{fig:nevllp}, where the area for which at 
least three events are produced is bounded with a red line.
The accessible mass range extends up to approximately 1.5~GeV, over a
range of $\epsilon$ between a few $10^{-5}$ and a few $10^{-4}$,
dependent on $m_{\dpho}$, beyond the sensitivity of the prompt analysis.
As explained in Section~\ref{sec:model},
when $m_{\dpho}$ is near the masses of the $\omega$ and $\phi$ resonances,
the \dpho~ decays dominantly into hadrons, and it has a shorter lifetime.
As a result the number of long-lived $A^{\prime}\rightarrow\mu^+\mu^-$
decays is suppressed, generating the discontinuities in the area
of expected sensitivity which can be seen in Figure~\ref{fig:nevllp}.

The next step in the analysis is a Monte Carlo study to verify whether 
the produced long-lived dark photons can be separated from the SM 
background using the proposed  IDEA detector. 

The events are first requested to pass the same initial preselection as for the prompt
analysis.

Reducible backgrounds from $e^+e^-\rightarrow\gamma\tau^+\tau^-$ with 
each $\tau$ decaying into $\mu\nu\bar{\nu}$ and $e^+e^-\rightarrow\gamma b\bar{b}$,
could give rise to final state signatures with a photon and a $\mu^+\mu^-$
pair detached from the interaction vertex. No event from the generated
Monte Carlo samples for these processes passes the kinematic preselections
and has an invariant mass in the interval 0.4-1.6~GeV,
except a single $\tau^+\tau^-$ event, which is removed by the vertex cleaning
cuts described below. No explicit simulation was performed for 
$e^+e^-\rightarrow\gamma c\bar{c}$ which is expected to be smaller than $e^+e^-\rightarrow\gamma b\bar{b}$.
These backgrounds are considered negligible for the following discussion. 

The irreducible prompt background $e^+e^-\rightarrow\gamma\mu^+\mu^-$
can be suppressed by applying a lower cut the transverse distance of the reconstructed vertex 
$r_{vx}$ from the centre of the detector. The variable in the plane transverse 
to the beam is used rather than the three dimensional variable $d_{vx}$, because
the position of the interaction vertex, for the present design of the FCC-ee collider \cite{FCC:2025uan}, has a spread of $\sim~0.4$~mm in the $z$ coordinate, as opposed to the $x$ and $y$ 
coordinates which have spreads below a few microns. 

A lower limit on $r_{vx}$ should reduce to zero the number of 
prompt background events. The key performance figure is in this case 
the quality of the reconstruction of  $r_{vx}$ in the IDEA tracker
for a vertex with two collimated high momentum muon tracks, 
which determines how low the threshold on $r_{vx}$ can be.
%Based on the parametrised simulation of the 
%IDEA inner tracker implemented in DELPHES, a detailed study was performed of 
%the factors affecting the measurement of $r_{vx}$ by comparing the true and 
%the measured value of $r_{vx}$ in the low mass signal events.

\begin{figure}[h] \centering
\includegraphics[width=0.45\textwidth]{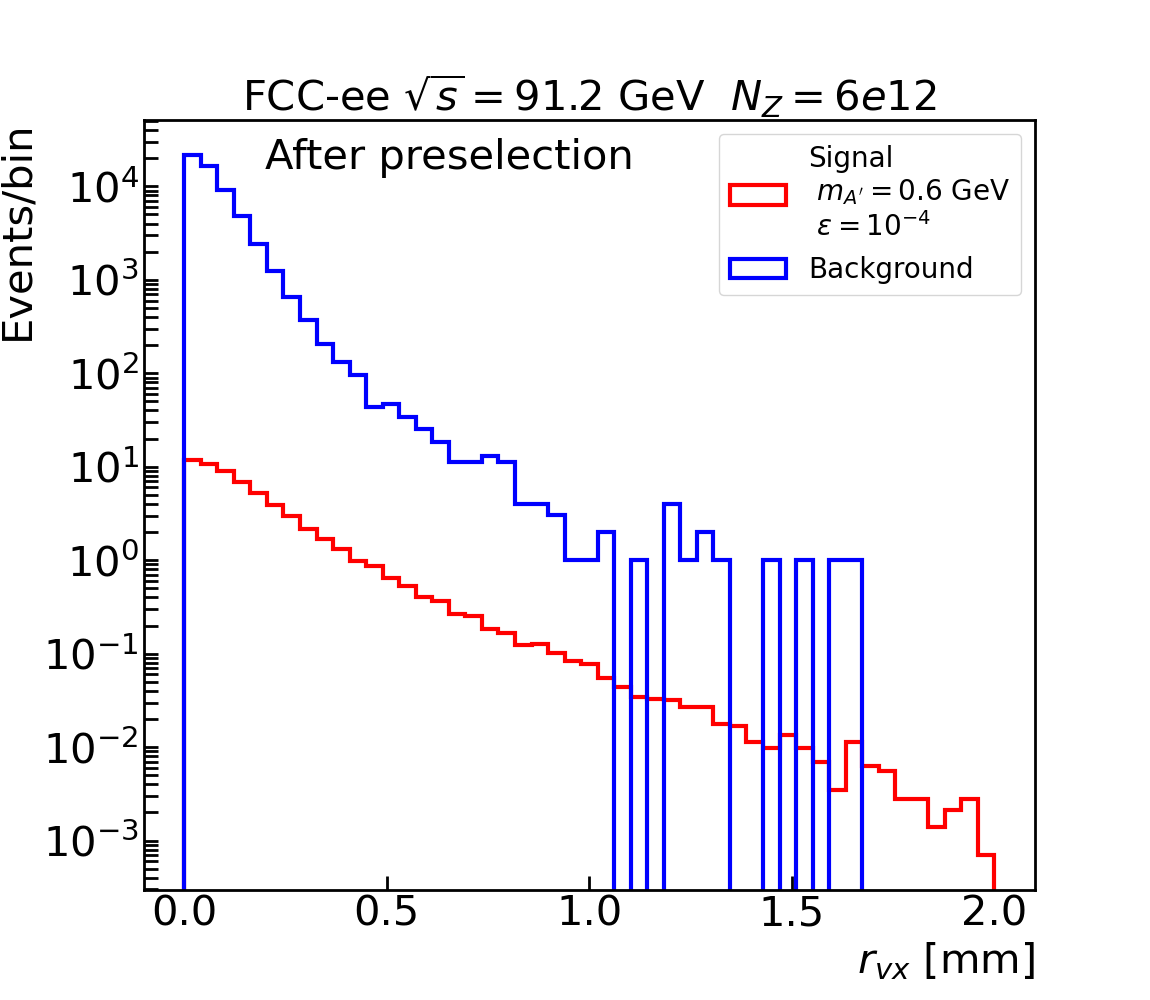}
\includegraphics[width=0.45\textwidth]{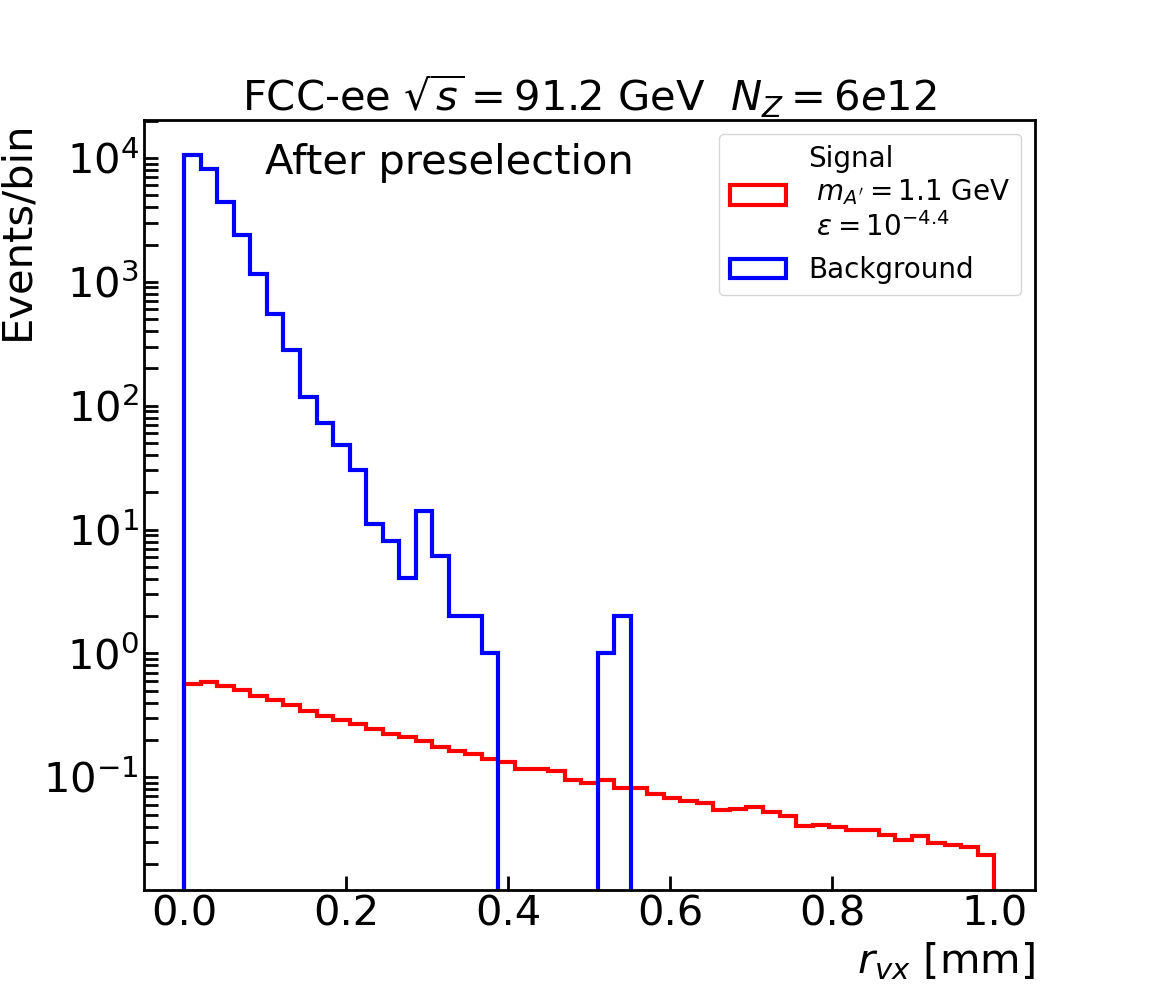}
	\caption{Distribution of $r_{vx}$ for signal and background for two 
	values of $m_{\dpho}$, 0.6 (left) and 1.1~GeV (right) after
	the preselection cuts. The events
	are selected in $m_{\mu^+\mu^-}$  windows in of 10~MeV around the 
	the nominal $m_{\dpho}$, and are normalised to the expected FCC-ee
	$Z$-pole run statistics.
        }\label{fig:rvxbefore}
\end{figure}

The distributions of $r_{vx}$ in the mass windows of $m_{\mu^+\mu^-}$ around 0.6 and 1.1~GeV
are shown in Figure~\ref{fig:rvxbefore} for signal and prompt background.
The values of $m_{\dpho}$ and $\epsilon$ for the signal are chosen to match the 
points with maximum number of signal events in Figure~\ref{fig:nevllp}.

Large non-gaussian tails towards high values of $r_{vx}$ are observed for the background, 
whereas the distribution for the signal is determined by the expected flight path of the 
\dpho. The effect is particularly clear for $m_{\mu^+\mu^-}~0.6$~GeV. In this
situation, the cut on $r_{vx}$ necessary to suppress the background level 
to a few events would also reduce the number of signal event below the detectable level.

The tails in the background $r_{vx}$ are from events where the detector fails
to reconstruct correctly the position of the vertex formed by the tracks of
the muon pair in the event.

\begin{figure}
\centering
\includegraphics[width=0.45\textwidth]{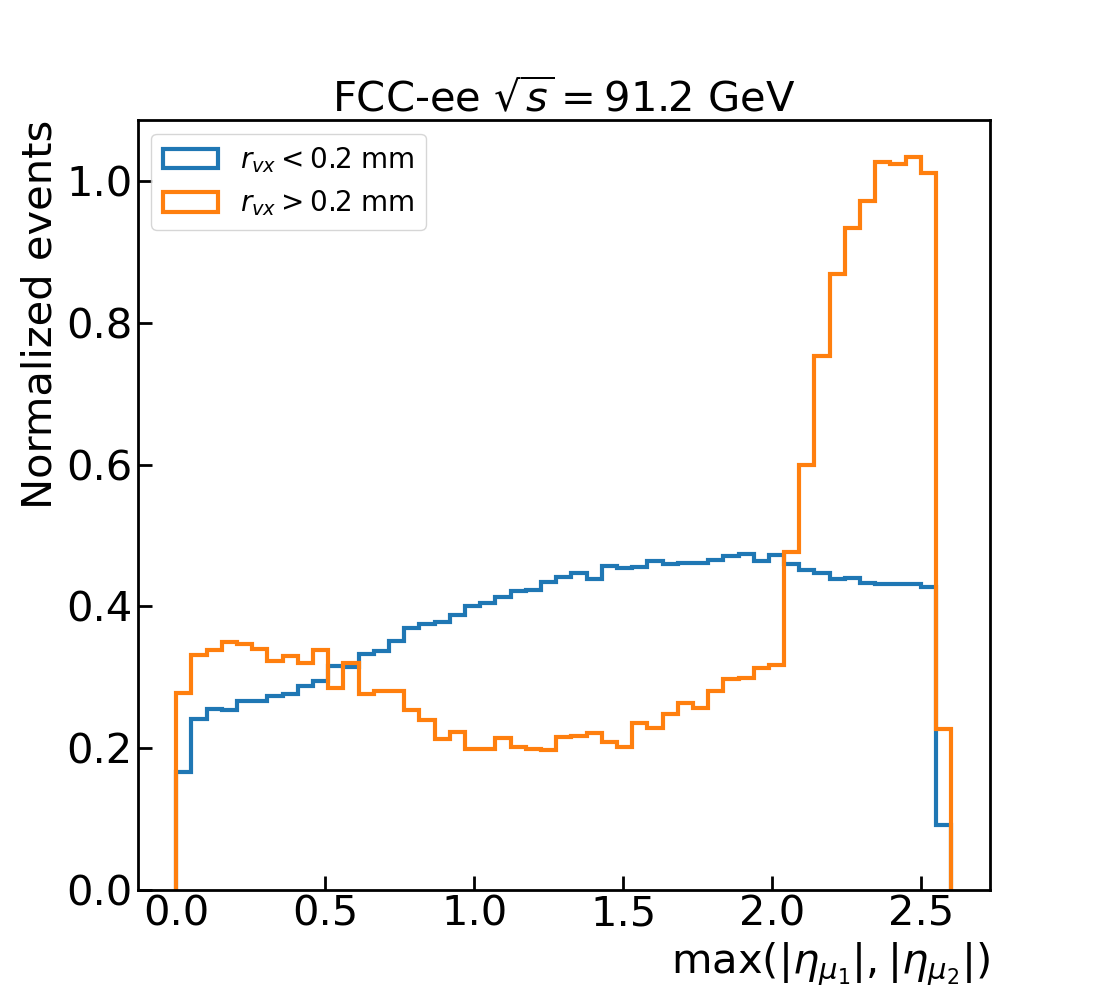}
\includegraphics[width=0.45\textwidth]{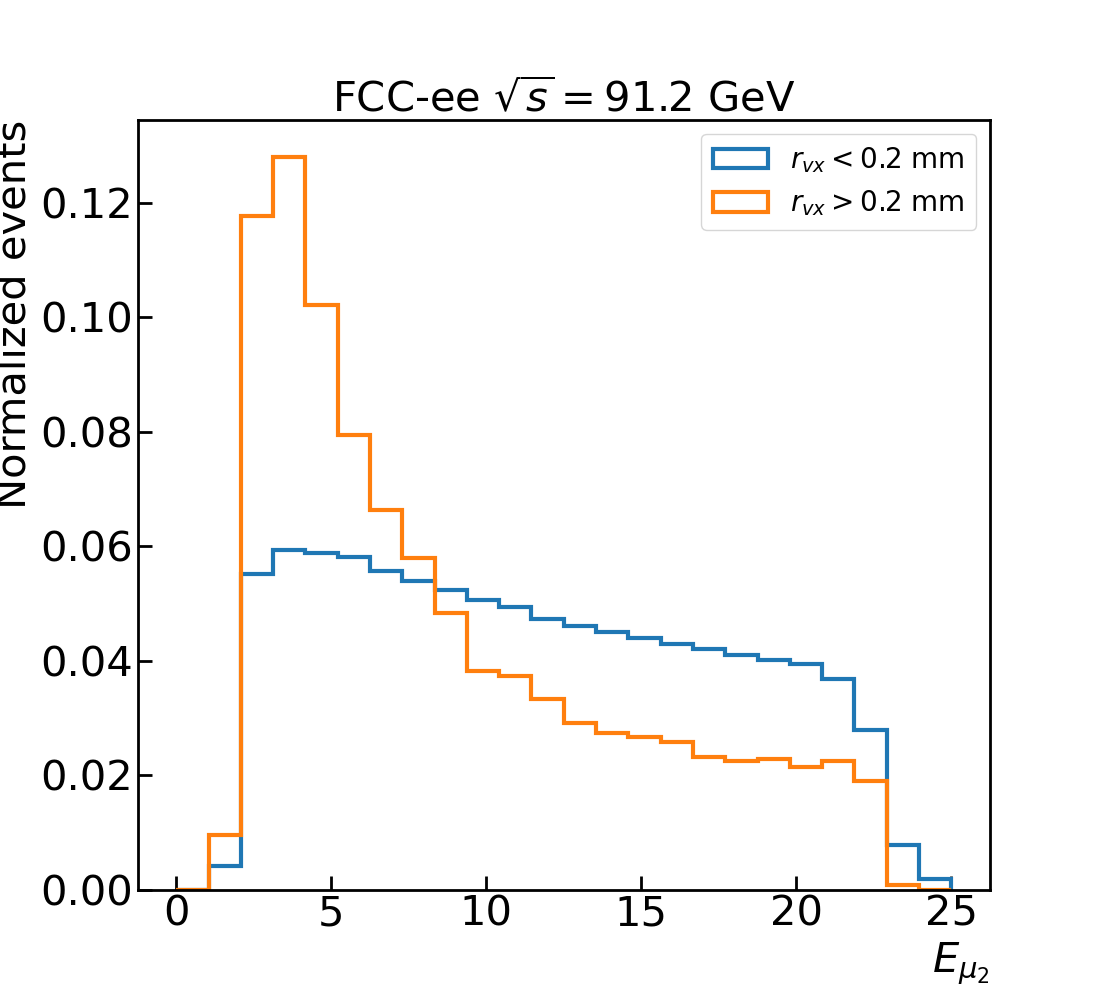}
	\caption{Distribution of the variables: left: max($|\eta_{\mu_1}|,|\eta_{\mu_2}|$),
	right: $E_{\mu_2}$ for $r_{vx}>0.2$~mm (orange) and $r_{vx}<0.2$~mm (blue)
	for the background $e^+e^-\rightarrow\gamma\mu^+\mu^-$ and $0.4<m_{\mu^+\mu^-}<2$~GeV.
         } \label{fig:maxeta}
\end{figure}

It is thus necessary to reject events which have a high probability that $r_{vx}$ is badly measured. Useful variables to perform this selection were found to be the energy of the softer muon $E_{\mu_2}$ and the maximum pseudorapidity of the two muons max($|\eta_{\mu_1}|,|\eta_{\mu_2}|$). 
Events where $r_{vx}$ is badly measured have
a low value of $E_{\mu_2}$ and a high value of max($|\eta_{\mu_1}|,|\eta_{\mu_2}|$).
This is shown in Figure~\ref{fig:maxeta}, where the two variables 
are plotted for the prompt background in the mass range of interest of the LLP analysis $0.4<m_{\mu^+\mu^-}<2$~GeV. 
For this background the truth value of $r_{vx}^{true}$ is determined  by the beam transverse spread and is smaller than 2.5~$\mu$m, therefore events  with $r_{vx}>0.2$~mm are badly measured. The selections max($|\eta_{\mu_1}|,|\eta_{\mu_2}|$)$<2$ and $E_{\mu_2}>6$~GeV strongly reduce background events with $r_{vx}>0.2$~mm.
%These cuts are also very powerful for reducing the tails in the measured $r_{vx}$ distribution, which are very important for a low-background search.

Another useful variable is $\Delta\phi_{vx\dpho}$,  
the angular distance in the transverse plane between the \dpho~ candidate and 
the vector connecting the centre of the detector and the position of the
reconstructed vertex.  In the case where the $x$ and $y$ coordinates of the real 
muon-muon vertex are equal to zero, the reconstructed vertex
can be either in the same hemisphere of the two reconstructed muons or in the opposite
hemisphere, with approximately equal probability. In the case where the real two-muon vertex
is detached from the beam direction, $\cos\Delta\phi_{vx\dpho}$ is predominantly positive.
A requirement $\cos\Delta\phi_{vx\dpho}>0.95$ reduces by a factor two the background with badly measured $r_{vx}$, and has approximately 100\% efficiency on the signal.
The combined requirements  max($|\eta_{\mu_1}|,|\eta_{\mu_2}|$), $E_{\mu_2}$ and $\cos\Delta\phi_{vx\dpho}$ will be collectively called `vertex cleaning cuts' in the following.

\begin{figure}[h] \centering
\includegraphics[width=0.45\textwidth]{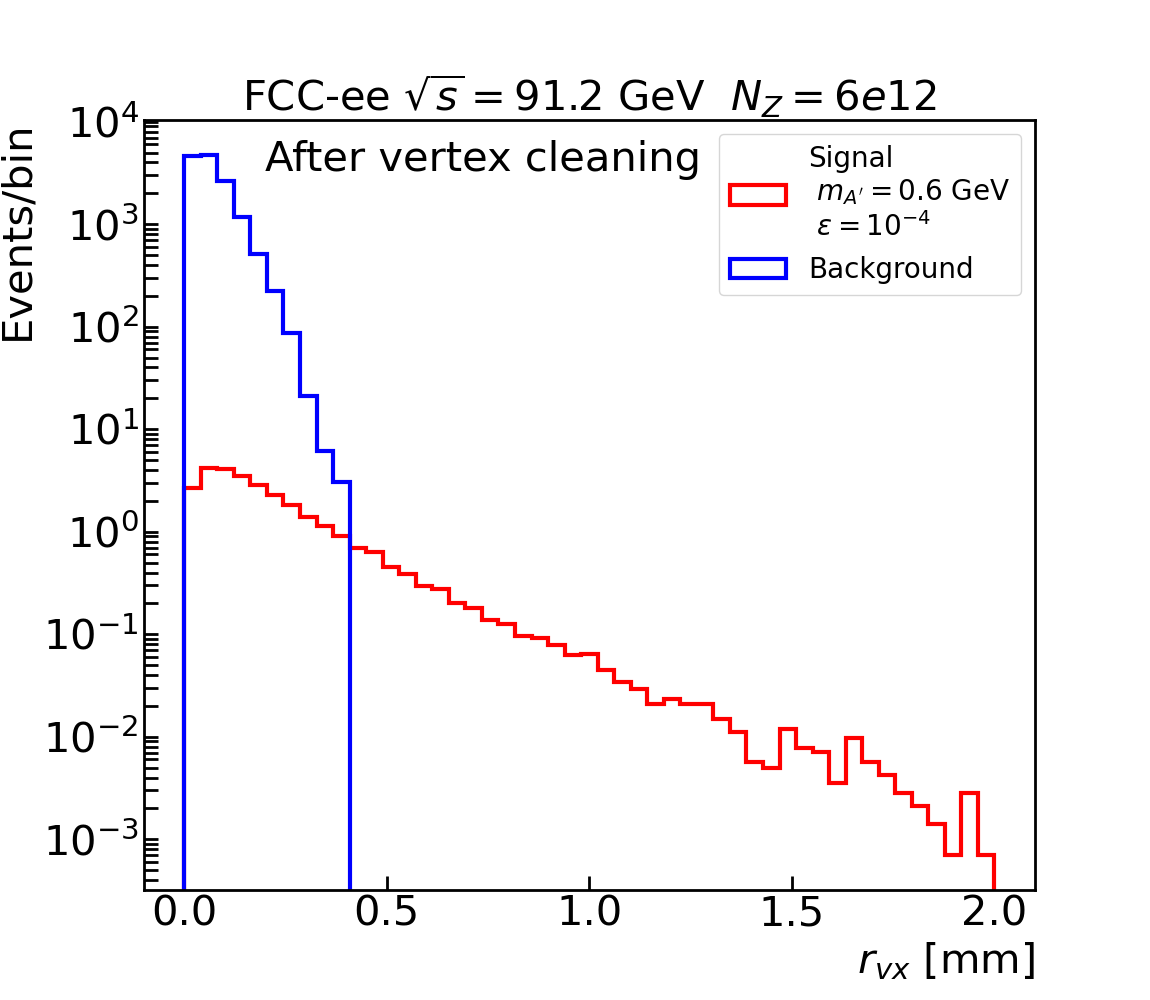}
\includegraphics[width=0.45\textwidth]{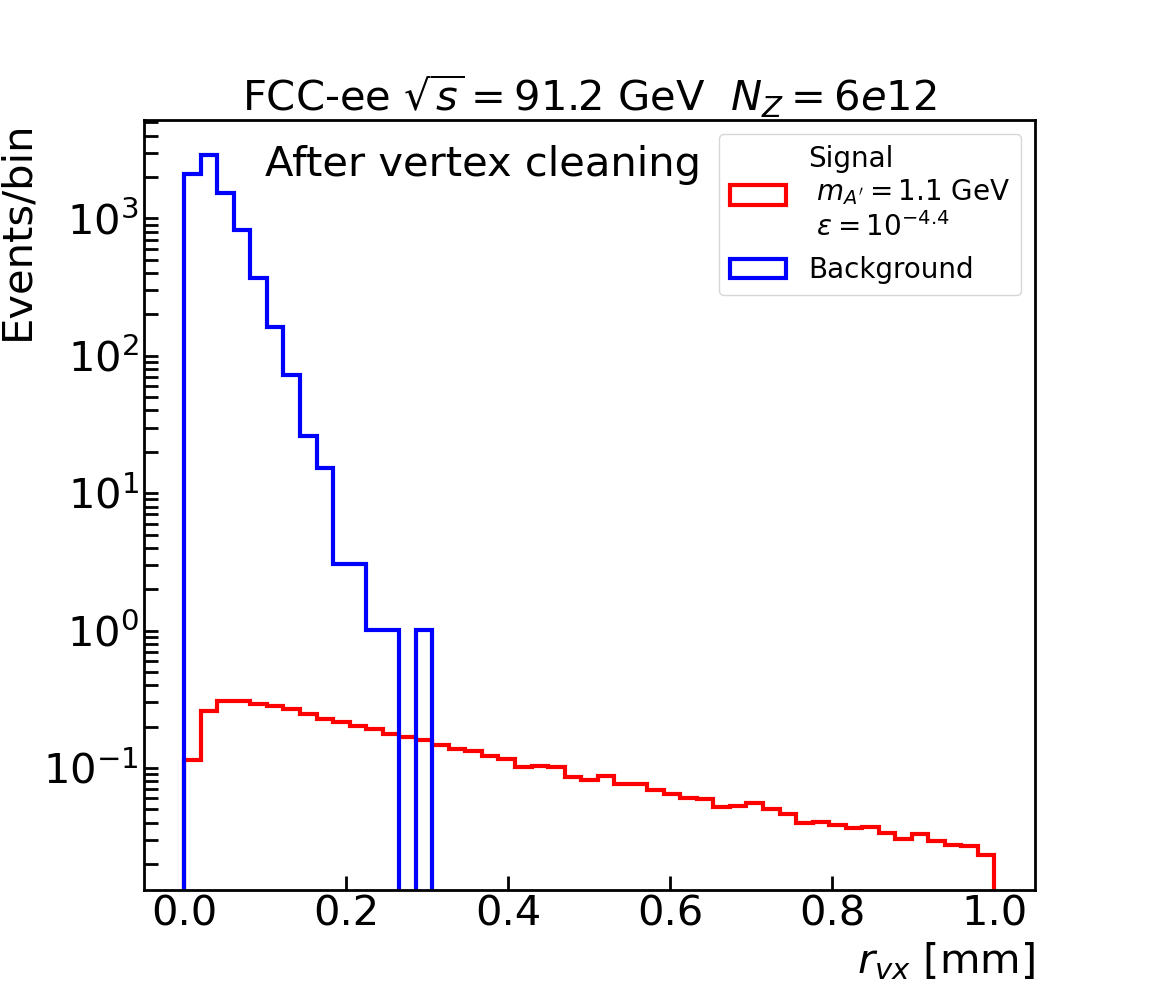}
        \caption{Distribution of $r_{vx}$ for signal and background for two
        values of $m_{\dpho}$, 0.6 (left) and 1.1~GeV (right) after
        the vertex cleaning  cuts. The events
        are selected in $m_{\mu^+\mu^-}$  windows in of 10~MeV around the
        the nominal $m_{\dpho}$, and are normalised to the expected FCC-ee
        $Z$-pole run statistics.
        }\label{fig:rvxafter}
\end{figure}

The impact of the vertex cleaning cuts can be observed in Figure~\ref{fig:rvxafter},
where the $r_{vx}$ distributions are plotted for the same signal and background 
samples as for Figure~\ref{fig:rvxbefore}
The badly measured events generating high values of $r_{vx}$ are effectively
removed, with a small reduction on the signal efficiency for high $r_{vx}$. 
For the final sensitivity, it is necessary to define as a 
function of $m_{\mu^+\mu^-}$ the optimal cut on $r_{vx}$, 
based on the resolution on the $r_{vx}$ measurement $\sigma_{r_{vx}}$.
A detailed study on $\sigma_{r_{vx}}$ was performed  using the parametrised simulation of the
IDEA tracker.

\begin{figure}
\centering
\includegraphics[width=0.49\textwidth]{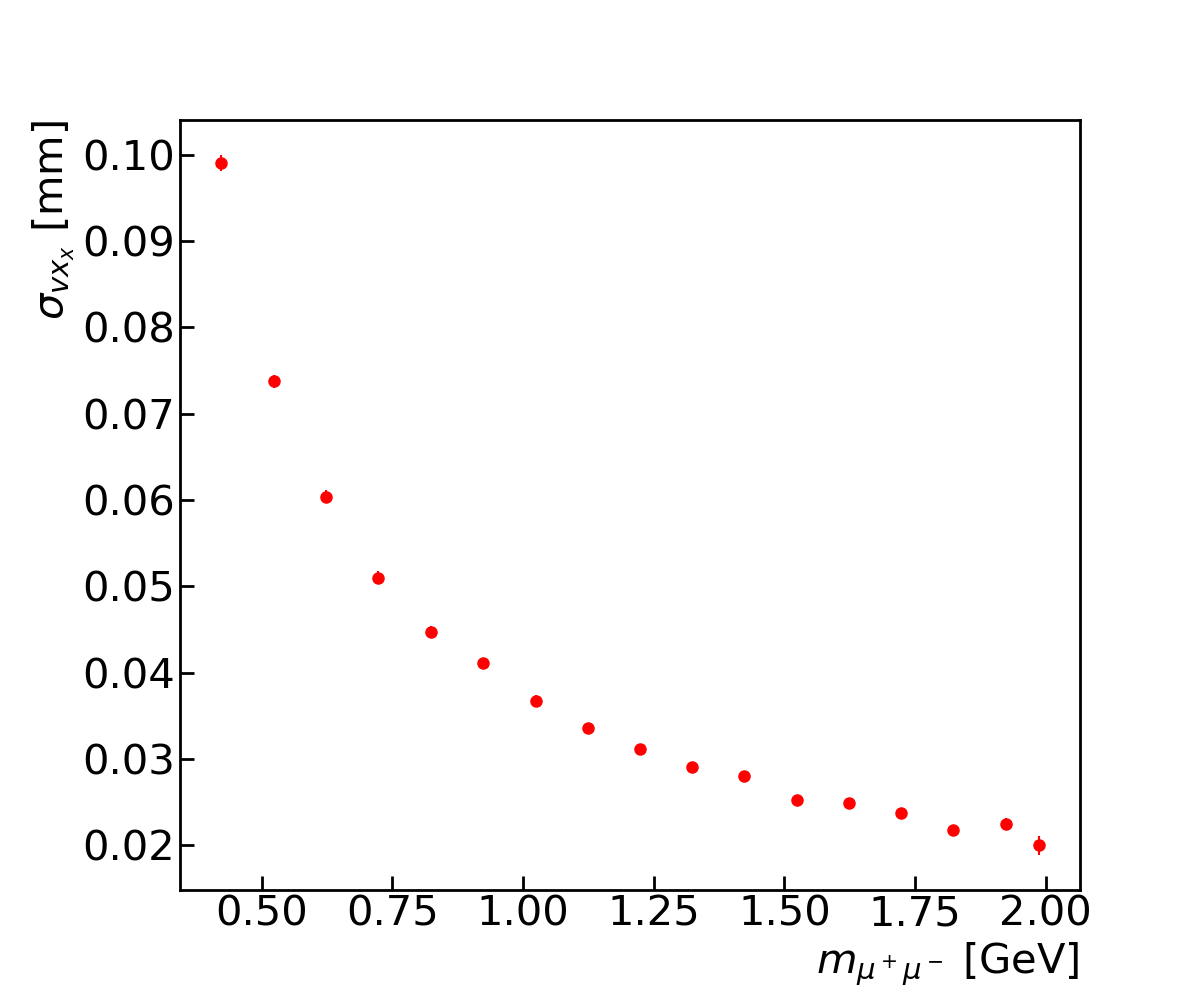}
\includegraphics[width=0.49\textwidth]{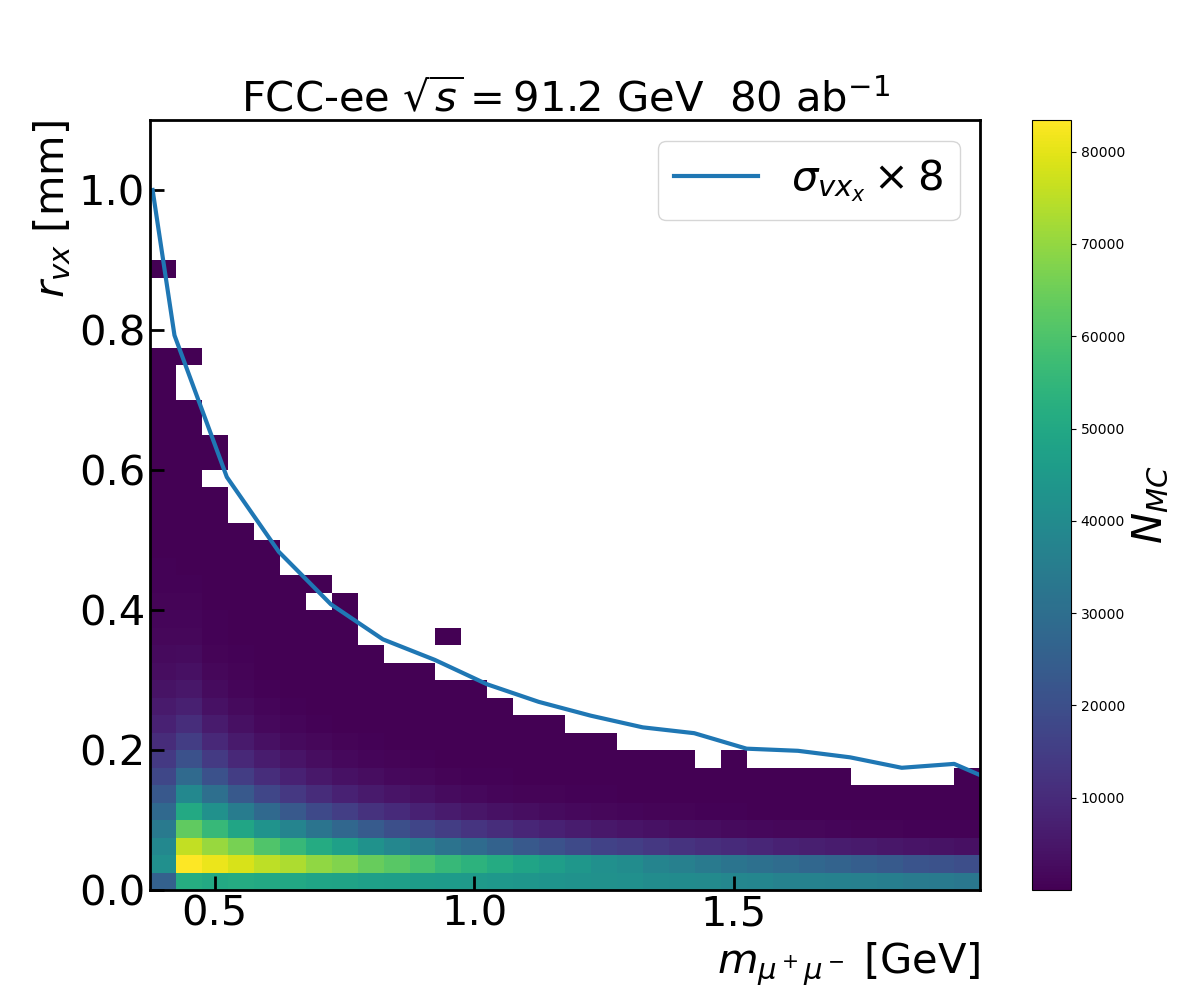}
        \caption{Left: Resolution on the $x$ component of the vertex position as a function 
	of $m_{\mu^+\mu^-}$ after the vertex selections described in the text, 
        for the background $e^+e^-\rightarrow\gamma\mu^+\mu^-$. Right: 
	number of MC background events passing in the $m_{\mu^+\mu^-}-r_{vx}$ plane.
        Superimposed in blue is the curve corresponding to eight times the resolution
	on $vx_x$.
         } \label{fig:resvert}
\end{figure}

The dependence of the resolution on the $x$ component of the vertex position 
on $m_{\mu^+\mu^-}$ for the prompt background after the vertex cleaning cut 
is shown in the right panel of  Figure~\ref{fig:resvert}.
The resolution $\sigma_{vx_x}$ is $\sim100~\mu$~mm for $m_{\mu^+\mu^-}=0.4$~GeV, and it becomes  $\sim20~\mu$m for $m_{\mu^+\mu^-}=2$~GeV. If the number of MC background events passing the vertex selections is plotted in the $m_{\mu^+\mu^-}-r_{vx}$ plane,
practically no events have $r_{vx}$ exceeding the calculated $\sigma_{vx_x}$  distribution multiplied by 8. This is shown in the left panel of Figure~\ref{fig:resvert}.

On this basis, the long-lived $\dpho\rightarrow\mu^+\mu^-$ candidates are
selected by requiring $r_{vx}>8\times\sigma_{vx_x}(m_{\mu^+\mu^-})$.

\begin{figure}
\centering
\includegraphics[width=0.75\textwidth]{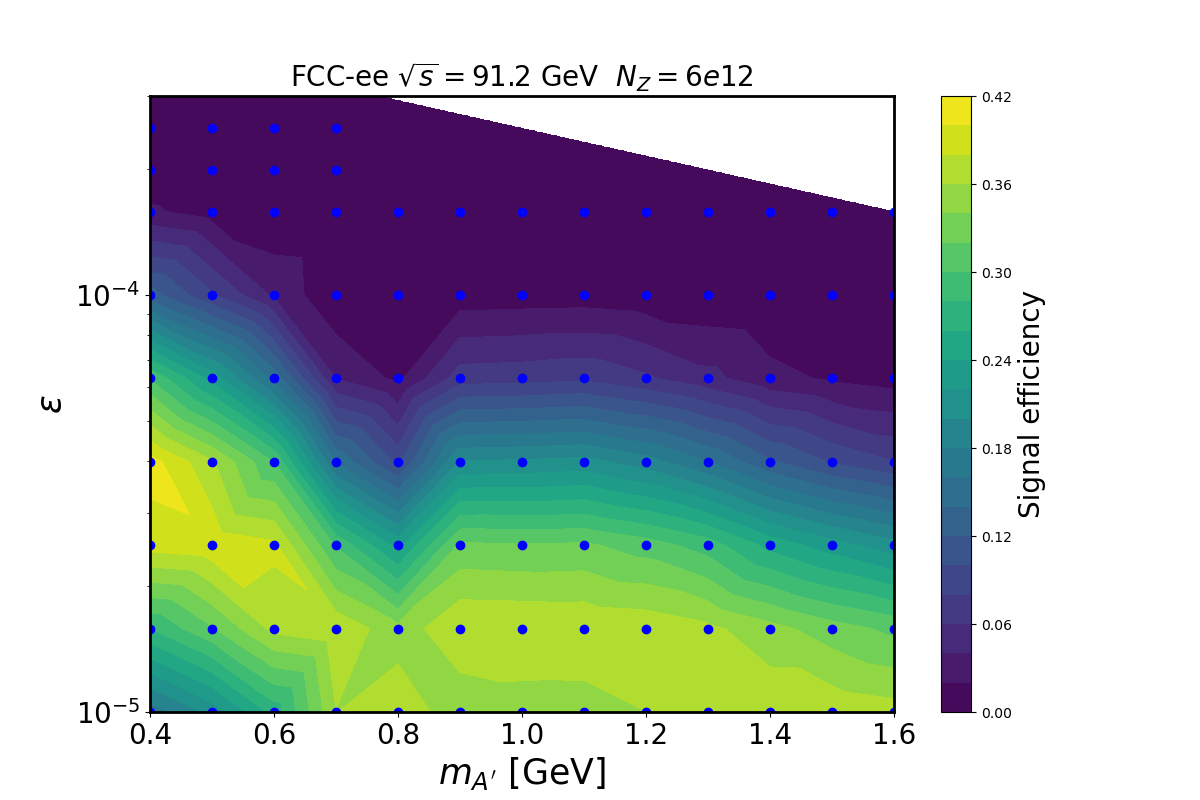}
        \caption{Signal efficiency for the long-lived selection
	described in the text in the $m_{\dpho}-\epsilon$ plane.
         } \label{fig:llpeff}
\end{figure}

Having thus defined a selection criteria for long-lived particles which 
minimises prompt background, the reach for the discovery of a
long-lived \dpho can be evaluated. For each generated value of $m_{\dpho}$ and $\epsilon$
the signal efficiency is calculated by applying the vertex-related cuts and by requiring 
that $m_{\mu^+\mu^-}$ is in a window of $\pm0.01$~GeV around $m_{\dpho}$. 
The resulting signal efficiency in the $m_{\dpho}-\epsilon$ plane is shown 
in Figure~\ref{fig:llpeff}.
The maximum efficiency is approximately 40\%, with a strong dependence on the mixing parameter $\epsilon$,
which determines the fraction of the events decaying in the IDEA inner detector that satisfy the selection on $r_{vx}$.

The background estimate needs to compensate for the statistical fluctuations of the number of
background events in the very narrow mass window, and for the fact that the
available MC statistics is  40\% of the expected statistics for the FCC-ee
Z-pole run. A total of 5 MC background events in the mass range 0.4-1.6~GeV 
pass the long-lived selection. Assuming that they are equally distributed among each of the 0.02~GeV width mass bins, the estimated average number of background events per mass bin, scaled to the statistics of the FCC-ee $Z$ run, is 0.21.

\begin{figure}
\centering
\includegraphics[width=0.7\textwidth]{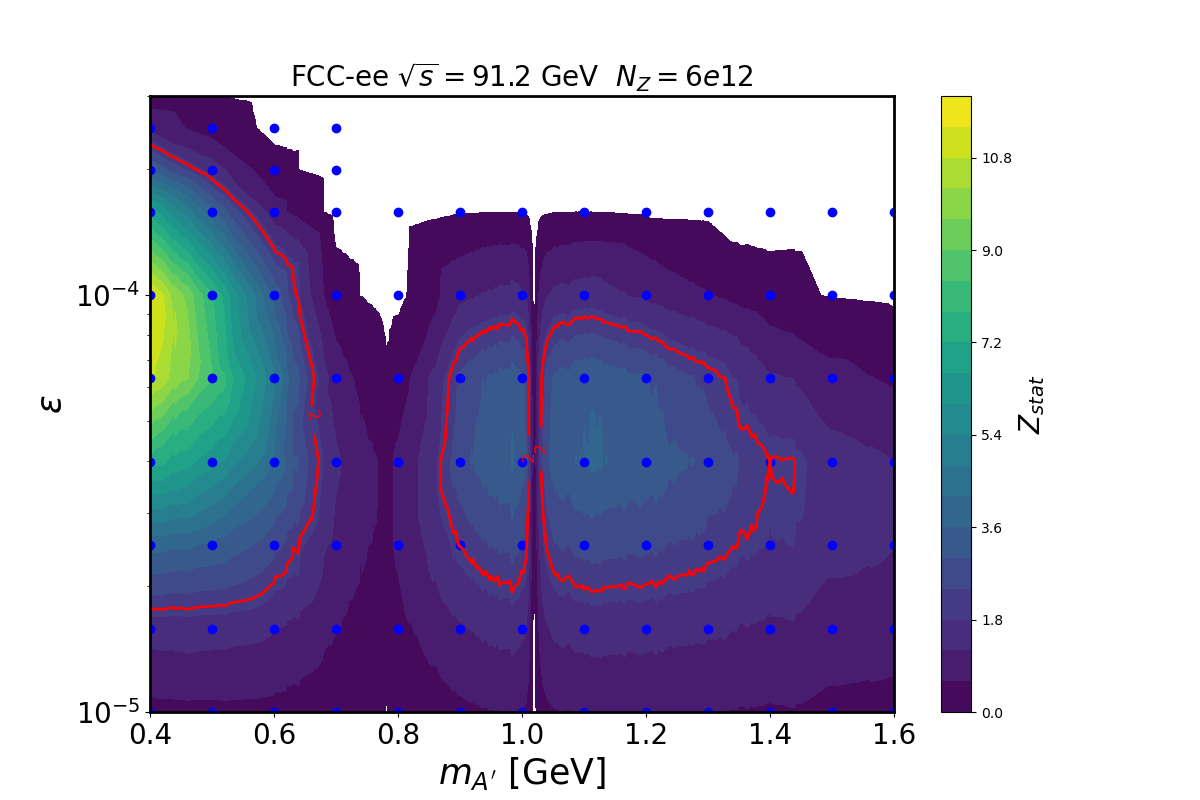}
	\caption{Statistical significance $Z_{stat}$ for the \dpho~ long-lived search
	in the $m_{\dpho}-\epsilon$ plane. 
	The lines bounding the region where $Z_{stat}>2$ are shown in red.
         } \label{fig:llpres}
\end{figure}

The final sensitivity ($Z_{stat}$) is calculated for each generated signal point from the
estimated number of signal and background events normalised to 
the expected statistics of the FCC-ee $Z$-pole run, based on the prescriptions in 
\cite{ATLASstat}. The results are  shown in Figure~\ref{fig:llpres}. 
The lines bounding the region where $Z_{stat}>2$ are shown in red, corresponding to
the 95\% CL sensitivity for $\dpho$.

\FloatBarrier 

\section{Results and Conclusions}\label{sec::conclusions}
\begin{figure} \centering
\includegraphics[width=0.7\textwidth]{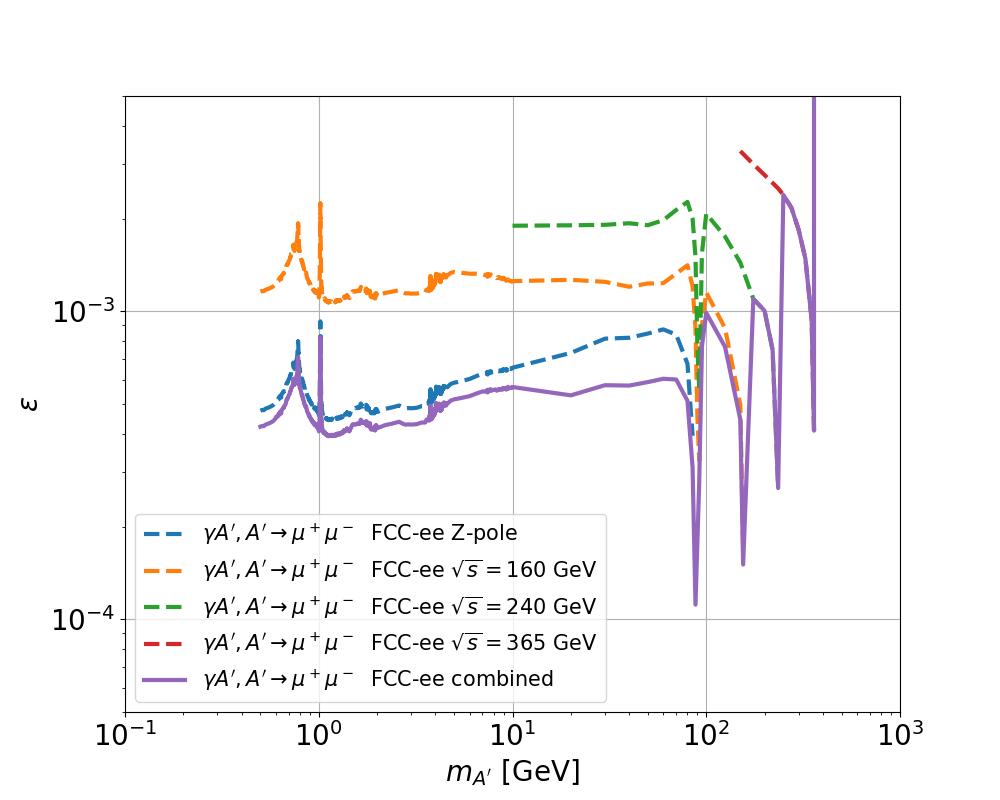}
\caption{Sensitivity limits at 95\% confidence level in the $m_{\dpho}-\epsilon$ plane
        for the prompt analysis. The results 
        for the different FCC-ee runs and for their statistical combination
	are shown.}\label{fig:dphoreach1}
\end{figure}

On the basis of the prompt analysis described in Section~\ref{sec::prompt},
the 95\% confidence level sensitivity limits on the $\epsilon$ parameter 
were calculated for all of the four different $\sqrt{s}$ runs of the FCC-ee.
The results are shown in Figure~\ref{fig:dphoreach1}.  

The limits cover a mass range 0.5-360~GeV.
The contributions of the different runs are statistically combined to yield 
the complete expected reach of FCC-ee running. In the mass region 10-85~GeV the 
$Z$-pole run yields the best sensitivity, as expected, but a significant  improvement
in reach is expected when combining the results with the data-taking at $\sqrt{s}=160$ and 240~GeV.

\begin{figure} \centering
\includegraphics[width=0.7\textwidth]{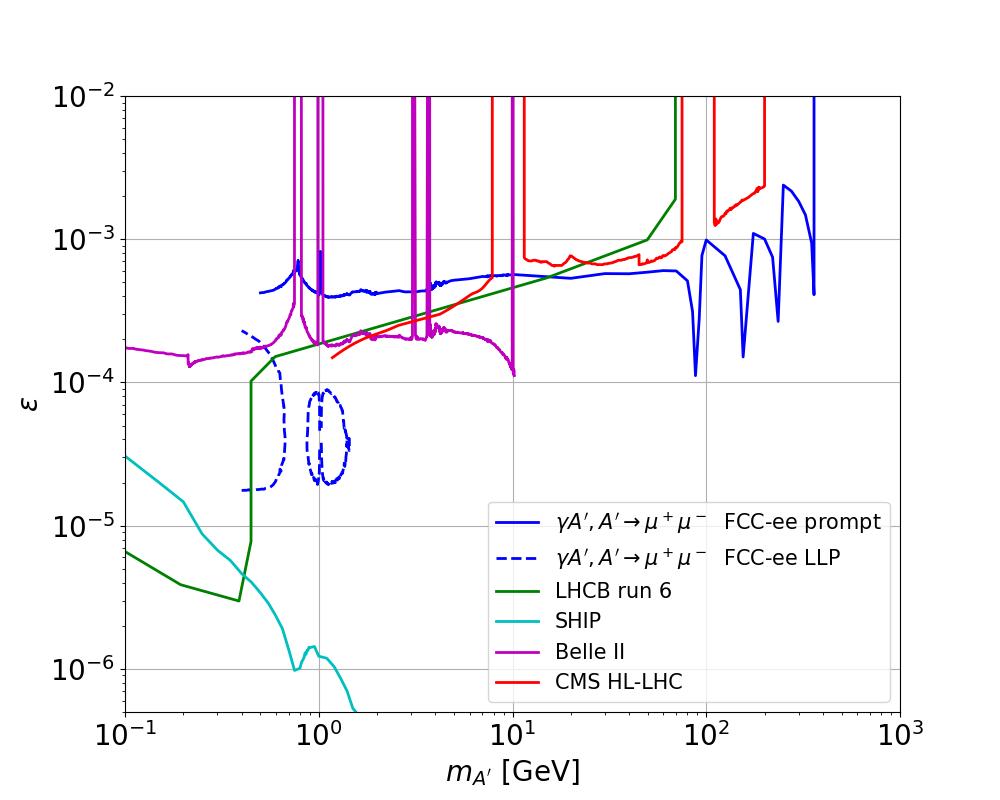}
\caption{Sensitivity limits at 95\% confidence level in the $m_{\dpho}-\epsilon$ plane
        for the statistical combination of the FCC-ee runs compared to the HL-LHC extrapolations for CMS \cite{CMS:2019buh,CMS:2023hwl} and LHCB \cite{Craik:2022riw} and to the projections for Belle II  \cite{Kou:2018nap} and SHIP \cite{SHiP:2020vbd}.}\label{fig:dphoreach}
\end{figure}

The combined FCC-ee reach for both prompt and long-lived analysis is compared with the extrapolations for high-luminosity LHC (HL-LHC)
for the CMS the LHCB experiments, and with the projected reach of the
SHIP and Belle II experiments in Figure \ref{fig:dphoreach}. 
For the CMS experiment, the extrapolation is calculated by scaling the 
expected reach in $\epsilon^2$ for the published LHC Run 2
analyses \cite{CMS:2019buh,CMS:2023hwl} by the 
square root of the ratio between 3~ab$^{-1}$, the design 
integrated luminosity for HL-LHC, and the integrated luminosity
used for the analyses. For LHCB an extrapolation to HL-LHC
is found in \cite{Craik:2022riw}. For SHIP and Belle II the projected reach 
is provided in \cite{SHiP:2020vbd} and \cite{Kou:2018nap} respectively. 

In the mass interval 20-365~GeV, the expected FCC-ee sensitivity 
for the prompt analysis improves on the HL-LHC extrapolations.

\begin{figure} \centering
\includegraphics[width=0.7\textwidth]{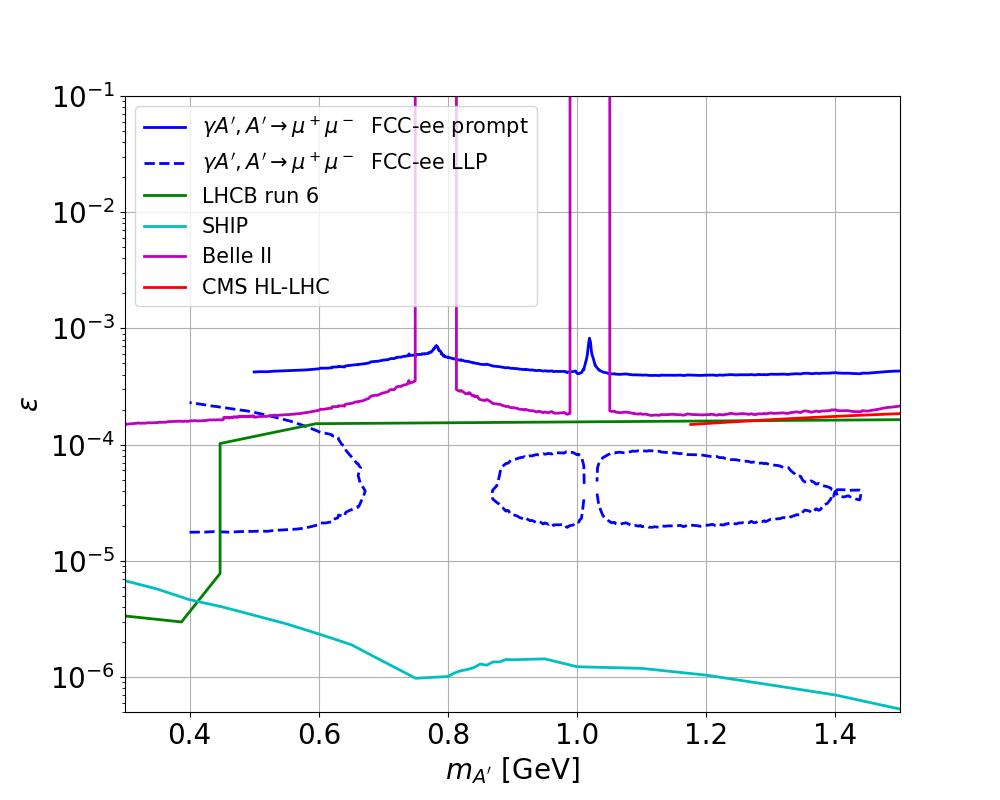}
\caption{Sensitivity limits at 95\% confidence level in the $m_{\dpho}-\epsilon$ plane
        for the statistical combination of the FCC-ee runs compared to the HL-LHC extrapolations for CMS \cite{CMS:2019buh,CMS:2023hwl} and LHCB \cite{Craik:2022riw} and to the projections for Belle II  \cite{Kou:2018nap} and SHIP \cite{SHiP:2020vbd}. Zoom on the sensitivity region of the LLP analysis.}\label{fig:dphoreachz}
\end{figure}

The long-lived analysis covers the mass range between 0.4 and 1.4 GeV in interval
of the mixing $\epsilon$ which is outside of the reach of the extrapolations of 
both the LHC and  the beam-dump experiments. This 
is shown more clearly in Figure~\ref{fig:dphoreachz}, where 
the relevant region in parameter space is shown with the mass in linear scale.
This results crucially depends on the capability of the vertexing and tracking system
of the experiment to measure with high precision the radial position of the
vertex built out of two very collimated tracks, and it constitutes a severe benchmark 
for the design of a future detector at the FCC-ee.

\FloatBarrier 

\bibliography{darkpho}% common bib file

\end{document}